\documentclass[11pt,letterpaper]{article}
\usepackage[T1]{fontenc}
\usepackage{lmodern}
\usepackage[letterpaper,margin=1in]{geometry}
\usepackage{amsmath,amssymb,amsthm,mathtools,booktabs,microtype}
\usepackage{enumitem}
\usepackage{tikz}
\usepackage[hidelinks]{hyperref}
\usepackage[nameinlink,noabbrev]{cleveref}
\allowdisplaybreaks
\setlist{itemsep=2pt,topsep=5pt}

\newtheorem{theorem}{Theorem}[section]
\newtheorem{lemma}[theorem]{Lemma}
\newtheorem{proposition}[theorem]{Proposition}
\newtheorem{corollary}[theorem]{Corollary}
\theoremstyle{definition}
\newtheorem{definition}[theorem]{Definition}
\newtheorem{example}[theorem]{Example}
\theoremstyle{remark}
\newtheorem{remark}[theorem]{Remark}
\numberwithin{equation}{section}
\crefname{theorem}{Theorem}{Theorems}
\crefname{lemma}{Lemma}{Lemmas}
\crefname{proposition}{Proposition}{Propositions}
\crefname{corollary}{Corollary}{Corollaries}
\crefname{definition}{Definition}{Definitions}
\crefname{example}{Example}{Examples}
\crefname{remark}{Remark}{Remarks}
\crefname{section}{Section}{Sections}
\crefname{subsection}{Section}{Sections}
\crefname{appendix}{Appendix}{Appendices}
\crefname{table}{Table}{Tables}
\crefname{figure}{Figure}{Figures}
\ifdefined\AddToHook
\AddToHook{env/lemma/begin}{\crefalias{theorem}{lemma}}
\AddToHook{env/proposition/begin}{\crefalias{theorem}{proposition}}
\AddToHook{env/corollary/begin}{\crefalias{theorem}{corollary}}
\AddToHook{env/definition/begin}{\crefalias{theorem}{definition}}
\AddToHook{env/example/begin}{\crefalias{theorem}{example}}
\AddToHook{env/remark/begin}{\crefalias{theorem}{remark}}
\fi

\newcommand{\E}{\mathbb E}
\newcommand{\R}{\mathbb R}
\newcommand{\Q}{\mathbb Q}
\newcommand{\Var}{\operatorname{Var}}
\newcommand{\TV}{\operatorname{TV}}
\newcommand{\poly}{\operatorname{poly}}
\newcommand{\clip}{\operatorname{clip}}
\newcommand{\dist}{\operatorname{dist}}
\newcommand{\depth}{\operatorname{depth}}
\newcommand{\ceil}[1]{\left\lceil #1\right\rceil}
\newcommand{\floor}[1]{\left\lfloor #1\right\rfloor}
\newcommand{\BIS}{\ensuremath{\#\mathrm{BIS}}}
\newcommand{\Hs}{\mathsf H}
\newcommand{\PL}{\Phi^{\mathrm L}}
\newcommand{\cP}{\mathcal P}
\newcommand{\Bd}{\mathcal B}
\newcommand{\ind}{\mathbf 1}

\title{Approximating \BIS\ in $\exp(\widetilde O(n^{1/3}))$ Time}

\author{Jingcheng Liu\thanks{School of Computer Science, Nanjing University, Nanjing, China.
Email: \href{mailto:liu@nju.edu.cn}{\texttt{liu@nju.edu.cn}}.}
\and Yitong Yin\thanks{School of Computer Science, Nanjing University, Nanjing, China.
Email: \href{mailto:yinyt@nju.edu.cn}{\texttt{yinyt@nju.edu.cn}}.}}
\hypersetup{pdftitle={Approximating \#BIS in exp(\textasciitilde O(n\textasciicircum(1/3))) Time}, pdfauthor={Jingcheng Liu, Yitong Yin}}
\date{}

\begin{document}
\hypersetup{pageanchor=false}
\begin{titlepage}
\maketitle
\thispagestyle{empty}
\begin{abstract}
\normalsize
We give randomized algorithms for approximately counting and nearly uniformly sampling independent sets in arbitrary bipartite graphs (\BIS).
At fixed accuracy, both algorithms run in $\exp(O(n^{1/3}\log^{4/3}n))$ time on $n$-vertex graphs.

More generally, the same running-time dependence on the number of variables holds for approximate counting and sampling with efficiently evaluable positive log-supermodular weights on Boolean variables.
A weighted Sinclair--Jerrum walk on the self-reduction tree reduces these tasks to coarse lower estimates of residual log-partition functions.
We obtain these estimates by optimizing a concave relaxation of an exact variational principle on the tree; threshold smoothing certifies that the low-degree search suffices.
\end{abstract}

\vspace{1.5em}
\setcounter{tocdepth}{1}
\begingroup
\footnotesize
\tableofcontents
\endgroup
\end{titlepage}
\hypersetup{pageanchor=true}
\pagenumbering{arabic}

\section{Introduction}\label{sec:intro}

Given a bipartite graph $G$, the problem \BIS\ asks for the number $Z(G)$ of its independent sets.
Exact counting is $\#\mathrm P$-complete even on bipartite graphs~\cite{PB83}.
In approximate counting, \BIS\ represents a broad class of problems equivalent to it under approximation-preserving reductions.
Dyer, Goldberg, Greenhill, and Jerrum~\cite{DGGJ04} showed that \BIS\ is complete for a logically defined class $\#\mathrm{RH}\Pi_1$ under these reductions.
Equivalent problems include counting downsets of a partial order and satisfying assignments of conjunctions of implications~\cite{DGGJ04}, counting stable matchings~\cite{IL86,DGGJ04,CGM12Stable}, and approximating the partition function of the ferromagnetic Ising model with mixed local fields~\cite{GJ07Fields}.
\BIS\ is also a benchmark for weighted constraints, spin systems, and list homomorphisms~\cite{BDGJM13,CDGJLMR15,GJ15Spin,GGJ17List}.

By the trichotomy of Dyer, Goldberg, and Jerrum~\cite{DGJ10}, every counting constraint satisfaction problem with a fixed finite set of Boolean relations is polynomial-time computable, \BIS-equivalent, or equivalent to counting satisfying assignments of Boolean formulas ($\#\mathrm{SAT}$).
Yet the approximation complexity of the \BIS-equivalent case remains unresolved: algorithmic techniques have not yielded a fully polynomial randomized approximation scheme (FPRAS), and hardness reductions have not established NP-hardness of approximation~\cite{DGGJ04}.
The problems in this equivalence class are conjectured to have \emph{intermediate approximation complexity}: they have no FPRAS, yet they are strictly easier to approximate than $\#\mathrm{SAT}$~\cite{DGGJ04,GJ12Potts,CGGGJSV16}.
Determining the approximation complexity of \BIS, and hence of this entire equivalence class, is a central open problem in the area of approximate counting.

\BIS\ also fits into the broader framework of \emph{log-supermodular} weights on Boolean variables: positive weights whose logarithms are supermodular set functions.
This is the FKG lattice condition~\cite{FKG71}, also called multivariate total positivity of order two (MTP$_2$)~\cite{KR80}.
The framework includes ferromagnetic Ising models with arbitrary local fields, whose partition functions are \BIS-equivalent~\cite{GJ07Fields}; when all fields favor the same spin, they admit an FPRAS~\cite{JerrumSinclair93Ising}.
It also allows terms coupling more than two variables.
For \BIS, summing out one side gives such a weight on the other, with no more variables than vertices.
Log-supermodularity also marks a complexity boundary for Boolean weighted counting CSPs with nonnegative constraints and all unary weights available: apart from a tractable class generated by unary weights and disequality, a non-log-supermodular constraint gives $\#\mathrm{SAT}$-equivalence~\cite[Theorem~10.2]{BDGJM13}.
Whether general log-supermodular counting reduces to \BIS\ remains open~\cite[Theorem~6]{CDGJLMR15}.

Substantial algorithmic progress has been made on restricted graph classes and in favorable parameter regimes.
\emph{High-temperature} algorithms exploit weak long-range dependence through correlation decay or rapid mixing.
The hard-core model weights an independent set $I$ by $\lambda^{|I|}$ for a fugacity $\lambda>0$.
On graphs of maximum degree at most a fixed $\Delta\ge3$, Weitz's correlation-decay algorithm~\cite{Wei06} gives a deterministic fully polynomial-time approximation scheme (FPTAS) whenever $\lambda<\lambda_c(\Delta)=(\Delta-1)^{\Delta-1}/(\Delta-2)^{\Delta}$, the uniqueness threshold of the infinite $\Delta$-regular tree.
At $\lambda=1$ this means maximum degree at most five, and Liu and Lu~\cite{LL15} need this bound on one side only.
With the same fugacity on both sides, Chen, Liu, and Yin~\cite{CLY23} show that $\lambda_c(\Delta)$ remains the uniqueness threshold when the degree bound $\Delta$ holds on one side only, and they sample rapidly below it.
\emph{Low-temperature} algorithms exploit concentration around structured phases.
Algorithmic Pirogov--Sinai theory gives high-fugacity algorithms on lattice regions with suitable boundary conditions~\cite{HPR20} and beyond lattices~\cite{CHP26}.
Polymer methods yield approximation algorithms on bounded-degree bipartite expanders~\cite{JKP20} and on bipartite graphs with sufficiently unbalanced degrees or fugacities~\cite{CP20}, the latter including $\lambda=1$.
On almost every random regular bipartite graph of sufficiently large fixed degree, recent work of Kocurek, Oveis Gharan, and Tjowasi~\cite{KOGT26}, together with~\cite{JKP20}, covers all fugacities.
Graph containers give an FPTAS for \BIS\ on weakly expanding regular bipartite graphs of large fixed degree, and inverse-polynomial relative approximation in subexponential time on regular bipartite graphs of growing degree~\cite{JPP23}; spectral methods give an FPRAS in the dense regular case~\cite{CDKP24}.
These guarantees rely on assumptions that need not hold on arbitrary bipartite graphs.

Recent announcements of subexponential-time algorithms for exactly counting satisfying assignments of fixed-width CNF formulas~\cite{OAI26,Claude26} would, if confirmed, yield $2^{O(n\log\log n/\sqrt{\log n})}$-time exact counting algorithms for \BIS\ on arbitrary $n$-vertex bipartite graphs.
Before these announcements, the best known approximation algorithm for \BIS\ was the deterministic approximation algorithm of Goldberg, Lapinskas, and Richerby~\cite{GLR21}, running in time $O(2^{0.2372n})\poly(\epsilon^{-1})$ for relative error $\epsilon$.
The conjectured intermediate approximation complexity of the \BIS-equivalent class motivates the search for algorithms substantially faster than those available for SAT.
Against the traditional exponential-time benchmark for SAT, obtaining a subexponential-time approximation algorithm for \BIS\ has been a prominent open direction in approximate counting (see, e.g.,~\cite{Perkins20,Liu23Lecture}).
The question was explicitly posed by Jenssen, Perkins, and Potukuchi~\cite[Question~1]{JPP23} and reiterated by Carlson, Davies, Kolla, and Potukuchi~\cite{CDKP24}.
If confirmed, the announced algorithms would answer the question as stated, while also changing the SAT benchmark.
The underlying algorithmic goal remains to exploit the special structure of \BIS\ to obtain algorithms substantially faster than those available for SAT.
Against the announced SAT bounds, a natural quantitative target is a fixed-power saving in the exponent: does \BIS\ admit a $2^{O(n^c)}$-time approximation algorithm at fixed accuracy for some fixed $0<c<1$?

\subsection{Results}\label{sec:results}
We resolve this question by giving randomized algorithms for approximately counting and nearly uniformly sampling independent sets in arbitrary bipartite graphs.
At fixed accuracy, both algorithms run in $\exp(O(n^{1/3}\log^{4/3}n))$ time on $n$-vertex graphs.
We state the precise bounds in the exact real-arithmetic model of \cref{sec:model}; implementations using bit operations and fair random bits have the same exponential factors (\cref{thm:finite}).

Write $\mathcal I(G)$ for the independent sets of $G$, $Z(G)=|\mathcal I(G)|$, and $\mathsf U_G$ for the uniform law on $\mathcal I(G)$.

\begin{theorem}[Bipartite independent sets]\label{thm:main}
There are randomized algorithms with the following guarantees for every bipartite graph $G$ with $n\ge2$ vertices.
\begin{enumerate}[label=(\roman*)]
\item
For every real $0<\eta\le1/2$, the sampling algorithm returns an independent set of $G$ whose law is within total variation distance $\eta$ of $\mathsf U_G$.
Its running time is at most
\[
\exp\!\left(O\left(n^{1/3}\log^{4/3}n\right)\right) \poly\left(n,\log(1/\eta)\right).
\]

\item
For all reals $\epsilon,\delta\in(0,1/2]$, the counting algorithm returns an estimate $\widehat Z$ satisfying
\[
\Pr\left[\left|\frac{\widehat Z}{Z(G)}-1\right|\le\epsilon\right]\ge1-\delta.
\]
Its running time is at most
\[
\exp\!\left(O\left(n^{1/3}\log^{4/3}n\right)\right) \poly\left(n,\epsilon^{-1},\log(1/\delta)\right).
\]
\end{enumerate}
Both bounds hold on every execution in the exact real-arithmetic model.
\end{theorem}

The constants are absolute.
For rational accuracy and failure-probability parameters, \cref{thm:finite} gives a bit implementation with polynomial overhead in their encoding lengths.

\paragraph{Log-supermodular weights.}
The algorithms follow from a general result for log-supermodular weights on Boolean variables.
A positive weight $W:\{0,1\}^d\to\Q_{>0}$ is \emph{log-supermodular} if
\begin{equation}\label{wt:condition}
W(x\wedge y)\,W(x\vee y)\ge W(x)\,W(y) \qquad(x,y\in\{0,1\}^d),
\end{equation}
where $\wedge$ and $\vee$ are the coordinatewise minimum and maximum.

The algorithms receive the dimension $d\ge2$ and use the following representation of $W$.
\begin{itemize}
\item
(\emph{Exact evaluation}) The weight is given by an exact evaluator using at most $T_W$ operations per value; in the bit implementation, $T_W$ denotes the bit cost of returning an exact rational value.

\item
(\emph{Value length}) The algorithms receive a bound $B\ge1$ on the bit length of every value of $W$, where the bit length of a rational is the larger of those of its numerator and denominator.
\end{itemize}
Write $Z_W=\sum_xW(x)$ for the partition function and $\mu_W=W/Z_W$ for the target law.

\begin{samepage}
\begin{theorem}[Log-supermodular weights]\label{thm:lsm}
For every log-supermodular weight $W$ given in this representation, there are randomized algorithms with the following guarantees.
\begin{enumerate}[label=(\roman*)]
\item
For every real $0<\eta\le1/2$, the sampling algorithm returns a configuration whose law is within total variation distance $\eta$ of $\mu_W$.
Its running time is at most
\[
\exp\!\left(O\left(\min\left\{d,\, d^{1/3}\log^{2/3}d\,\log^{2/3}(d+B) \right\}\right)\right) \poly\left(d,B,T_W,\log(1/\eta)\right).
\]

\item
For all reals $\epsilon,\delta\in(0,1/2]$, the counting algorithm returns an estimate $\widehat Z$ satisfying
\[
\Pr\left[\left|\frac{\widehat Z}{Z_W}-1\right|\le\epsilon\right] \ge1-\delta.
\]
Its running time is at most
\[
\exp\!\left(O\left(\min\left\{d,\, d^{1/3}\log^{2/3}d\,\log^{2/3}(d+B) \right\}\right)\right) \poly\left(d,B,T_W,\epsilon^{-1},\log(1/\delta)\right).
\]
\end{enumerate}
Both bounds hold on every execution in the exact real-arithmetic model.
\end{theorem}
\end{samepage}

When $B$ and $T_W$ are polynomial in $d$, both algorithms run in $\exp(O(d^{1/3}\log^{4/3}d))$ time at fixed accuracy.
The algorithms access $W$ only through its evaluator; no factorization into local interactions or correlation-decay assumption is required.

To apply \cref{thm:lsm} to \BIS, let $G=(L,R,E)$ be a bipartite graph and write $N(S)$ for the neighborhood of a vertex set $S$ and $N(r)=N(\{r\})$.
Summing out $R$ gives the positive weight
\begin{equation}\label{intro:shoreweight}
W(S)=2^{|R\setminus N(S)|}\qquad(S\subseteq L),
\end{equation}
since $S$ can be completed to an independent set by adjoining any subset of $R\setminus N(S)$.

\begin{proof}[Proof of \cref{thm:main} from \cref{thm:lsm}]
If $|L|\le1$, the at most two choices of $S\subseteq L$, together with~\eqref{intro:shoreweight}, give exact counting and sampling algorithms in polynomial time.
Otherwise set $d=|L|\ge2$.
For $S,S'\subseteq L$, we have $N(S\cup S')=N(S)\cup N(S')$ and $N(S\cap S')\subseteq N(S)\cap N(S')$, so
\[
|N(S\cup S')|+|N(S\cap S')| \le |N(S)|+|N(S')|.
\]
Thus the coverage function $S\mapsto|N(S)|$ is submodular, so $\log W$ is supermodular and $W$ satisfies~\eqref{wt:condition}.
The values are integers at most $2^n$, so we may take $B=n+1$ and $T_W=\poly(n)$.
Since $Z_W=Z(G)$, the counter of \cref{thm:lsm} counts independent sets.

For sampling, complete the sampler's output $S$ by adjoining a uniform subset of $R\setminus N(S)$, using at most $n$ additional uniform draws.
This maps $\mu_W$ to $\mathsf U_G$ without increasing total variation distance.
The running-time bounds follow from \cref{thm:lsm} with $d\le n$ and $B=n+1$.
\end{proof}

When $k=\min\{|L|,|R|\}\ge2$, choosing the smaller side as $L$ gives the exponent
\[
O\left(\min\left\{k,k^{1/3}\log^{2/3}k\,\log^{2/3}n\right\}\right).
\]

Both algorithms also apply under every feasible pinning, that is, one whose occupied vertices form an independent set.
Deleting the occupied vertices, their neighbors, and the pinned vacant vertices leaves a bipartite induced subgraph.
If this subgraph has fewer than two vertices, count and sample by exact enumeration.
Restoring the occupied vertices gives a bijection between its independent sets and those consistent with the pinning.

\paragraph{The bipartite hard-core model.}
With positive rational vertex fugacities $\lambda_v$, the hard-core model assigns each independent set $I$ weight $\prod_{v\in I}\lambda_v$.
Summing out $R$ gives
\[
W_\lambda(S)= \prod_{v\in S}\lambda_v \prod_{r\in R\setminus N(S)}(1+\lambda_r), \qquad S\subseteq L.
\]
Its partition function is the hard-core partition function of $G$.
Up to an additive constant, $\log W_\lambda$ is a modular function minus the weighted coverage function $S\mapsto\sum_{r\in N(S)}\log(1+\lambda_r)$, so $W_\lambda$ is log-supermodular.
Complete the sampler's output $S$ by occupying each $r\in R\setminus N(S)$ independently with probability $\lambda_r/(1+\lambda_r)$.
This completion does not increase total variation distance.

The weight is exactly evaluable in time polynomial in $n$ and the total bit length of the fugacities, with $B$ at most $n$ plus that length.
Thus, with exact enumeration of $S$ when $|L|\le1$, \cref{thm:lsm} covers the hard-core model on arbitrary bipartite graphs with arbitrary positive rational vertex fugacities.
When the fugacities' total bit length is polynomial in $n$, the algorithms preserve $\exp(O(n^{1/3}\log^{4/3}n))$ time at fixed accuracy.

Bipartiteness makes this marginalization produce a positive log-supermodular weight, although the original hard-core edge constraint $\ind\{x_ux_v=0\}$ fails~\eqref{wt:condition}.
For every fixed $\Delta\ge3$ and fixed rational $\lambda>\lambda_c(\Delta)=(\Delta-1)^{\Delta-1}/(\Delta-2)^\Delta$, approximating the partition function at uniform fugacity $\lambda$ is \BIS-hard on bipartite graphs of maximum degree $\Delta$~\cite[Corollary~3]{CGGGJSV16}.
On general $\Delta$-regular graphs at the same fugacity, there is a constant $c=c(\Delta,\lambda)>0$ such that approximation within a factor $e^{cn}$ is NP-hard under randomized reductions~\cite[p.~2387]{SS14}.

\paragraph{The random-cluster and ferromagnetic Potts models.}
Let $G=(V,E)$ be a graph, not necessarily bipartite, with $n=|V|$ vertices and $m$ edges, and for $A\subseteq E$ let $\kappa(A)$ be the number of connected components of $(V,A)$, isolated vertices included.
With a rational $q\ge1$ and positive rational edge weights $v_e$, the random-cluster model gives $A$ the weight
\[
W_{q,v}(A)=q^{\kappa(A)}\prod_{e\in A}v_e,\qquad A\subseteq E,
\]
and its partition function $Z_{W_{q,v}}$ is the multivariate Tutte polynomial $Z_G(q,v)$~\cite[(1.1)]{Sokal05}.
The rank $n-\kappa(A)$ of $A$ in the cycle matroid of $G$ is submodular, so $\log W_{q,v}$ is supermodular when $q\ge1$; this is the FKG lattice condition for the random-cluster measure~\cite{Grimmett06}.
Computing $\kappa(A)$ by graph search evaluates the weight exactly in time polynomial in $n$, $m$ and the bit lengths of $q$ and the $v_e$.
Since $\kappa(A)\le n$, we may take $B$ to be $n$ times the bit length of $q$ plus the total bit length of the $v_e$.
If $m\le1$, enumerating the at most two edge sets counts and samples exactly; otherwise \cref{thm:lsm} applies with $d=m$.
When the bit lengths are polynomial in $n+m$, the algorithms run in $\exp(O(m^{1/3}\log^{4/3}(n+m)))$ time at fixed accuracy: $\exp(O(n^{1/3}\log^{4/3}n))$ on graphs of bounded degree, and $\exp(O(n^{2/3}\log^{4/3}n))$ on all graphs.

Multigraphs reduce to simple graphs: a loop $e$ contributes $1+v_e$, while a parallel bundle $F$ becomes one edge of activity $\prod_{e\in F}(1+v_e)-1$.
Lift samples by including loops independently with probabilities $v_e/(1+v_e)$ and replacing each present bundle by a nonempty $S\subseteq F$ according to weights $\prod_{e\in S}v_e$.
The latter is the law of independent Bernoulli choices with probabilities $v_e/(1+v_e)$ conditioned on being nonempty.
Sequential conditioning, using suffix products of the exclusion probabilities, samples this law in $O(|F|)$ arithmetic operations and uniform draws.
For a matroid $M$ on $E$ with a polynomial-time rank oracle, the weight $q^{-r_M(A)}\prod_{e\in A}v_e$ is log-supermodular because $r_M$ is submodular, and sums to the multivariate Tutte polynomial $\tilde Z_M(q,v)$~\cite[(1.3)]{Sokal05}.
Use $d=|E|$, replace $n$ by $d$ in the bound on $B$, and enumerate directly when $d\le1$.

For an integer $q\ge2$ and rational $y_e>1$, the ferromagnetic $q$-state Potts model assigns a coloring $\sigma:V\to\{1,\ldots,q\}$ the weight $\prod_ey_e$ over the edges $e$ whose endpoints receive the same color.
Writing each factor as $1+(y_e-1)\ind\{\text{the endpoints of $e$ agree}\}$ and expanding gives the Fortuin--Kasteleyn representation: the Potts partition function equals $Z_G(q,v)$ with $v_e=y_e-1$~\cite[Theorem~2.3]{Sokal05}, so the counter estimates it.
Coloring each component of the sampler's output $A$ independently and uniformly from $\{1,\ldots,q\}$ maps the random-cluster law to the Potts law, by the Edwards--Sokal coupling~\cite{EdwardsSokal88}, and does not increase total variation distance.
For every integer $q\ge3$, approximating the ferromagnetic Potts partition function is \BIS-hard~\cite{GJ12Potts}, whereas the ferromagnetic Ising model, the case $q=2$, has an FPRAS~\cite{JerrumSinclair93Ising}.

\subsection{Overview of the proof}\label{sec:technicaloverview}

\paragraph{Self-reduction and coarse estimates (\cref{sec:coarseoracle}).}
Sampling and counting reduce to the conditional marginal probabilities along the \emph{self-reduction tree}.
Fix an order of the $d$ coordinates; the nodes are the prefixes $x_v\in\{0,1\}^j$, $0\le j\le d$, with children $v0$ and $v1$, and the \emph{mass} $Z_v$ of a node is the partition function of the \emph{residual} weight left by pinning its prefix.
The next coordinate is $1$ with conditional marginal probability $p_v=Z_{v1}/Z_v$.
The classical self-reduction as in~\cite{JVV86} multiplies estimated probabilities, so their errors accumulate along a path and it needs relative accuracy $O(1/d)$ at every node.
A different self-reduction by Sinclair and Jerrum~\cite[Section~4]{SJ89} uses a reversible tree walk and needs much less.
Because the leaves carry exact weights, its stationary law conditioned on the leaves is exactly the target law.
Mass estimates with approximation ratio $r=e^b$, which we call the \emph{distortion}, cost only a factor $e^{O(b)}$ in running time (\cref{conv:conversion}).
Our task is therefore to estimate the logarithm of every residual mass from below within an additive error $b$ as in~\cref{conv:oracle}; since pinning preserves log-supermodularity, the same estimator applies to every node along the self-reduction tree.

\paragraph{A variational principle on the self-reduction tree (\cref{sec:treeprinciple}).}
Lower estimates come from Gibbs' variational principle~\cite{WJ08}, which we write on the tree.
Express a strictly positive law as $\mu=e^\Phi/Z_\Phi$, with \emph{potential} $\Phi:\{0,1\}^d\to\R$ and $Z_\Phi=\sum_xe^{\Phi(x)}$.
A \emph{branching rule} assigns a probability $\theta_v$ to moving from each internal node $v$ to $v1$; write $\nu_\theta(v)$ for the probability of reaching $v$, and $\nu_\theta$ for the law of the leaf reached, which can be any law.
Write $\Hs(\nu)=-\sum_x\nu(x)\log\nu(x)$ for the entropy of a finite law $\nu$ and $\Hs(p)=-p\log p-(1-p)\log(1-p)$ for the binary entropy of $p\in[0,1]$, with $0\log0=0$.
By the chain rule for entropy, $\Hs(\nu_\theta)=\sum_v\nu_\theta(v)\Hs(\theta_v)$, where the sum runs over internal nodes.
Gibbs' principle therefore becomes
\[
\log Z_\Phi=\max_\theta\Bigl( \sum_v\nu_\theta(v)\Hs(\theta_v)+\E_{\nu_\theta}\Phi\Bigr),
\]
with the conditional marginal probabilities $\theta=p$ as its unique maximizer (\cref{lat:treeprinciple}).
Every branching rule thus gives a lower bound, and the best one is exactly what self-reduction needs.
The objective, however, is not concave in $\theta$, even for log-supermodular laws.
It is concave in the leaf probabilities, but evaluating its entropy requires those probabilities.

\paragraph{Relaxing to guides (\cref{sec:descents,sec:lovasz}).}
The above variational formulation is not friendly to optimization algorithms.
Both parts inside its objective stand in the way of concavity: the non-linear entropy terms, and the expected potential that depends on the branching probabilities through products along paths.
There is a standard way to linearize an entropy (see~\cref{lat:onebit,rem:hj}), and for the potential term we apply the Lov\'asz extension in~\cref{lat:lovasz}.
The identity~\eqref{lat:onebit} linearizes a binary entropy as $\Hs(p)=\E[a(U)\ind\{U\ge1-p\}]$ for $U$ uniform on $(0,1)$ and the logit $a(u)=\log(u/(1-u))$.
Driving the descent by independent uniform inputs $U_1,\ldots,U_d$, one per depth, with the binary decision rule $F_i=\ind\{U_i\ge1-\theta_v\}$ at the node $v$ reached at depth $i-1$, turns the entropy terms at depth $i-1$ into $\E[a(U_i)F_i]$.
The binary decision rules $\ind\{U\ge1-p\}$ do not form a convex domain (for optimization), so we have to further relax to an appropriate set of real-valued decision functions $F_i\in[0,1]$ of $U_1,\ldots,U_i$ that we call \emph{guides} (see~\cref{lat:guide}).
This relaxation also means that the potential term needs to be evaluated at fractional decisions, which is provided by its Lov\'asz extension $\PL$~\cite{Lov83} after normalizing by $\Phi(0)=0$.
This gives the \emph{variational objective} in~\cref{lat:score}
\[
\cP(F)=\sum_{i=1}^d\E[a(U_i)F_i(U)]+\E\PL(F(U)),
\]
a \emph{linear score} term plus a \emph{potential term}.
Every guide satisfies $\cP(F)\le\log Z_\Phi$, with equality only for the \emph{exact guide} $f$, the guide defined by the true conditional marginal probabilities.
Furthermore, $\cP$ is concave exactly when $\mu$ is log-supermodular (\cref{lat:variational}).
It is an expectation over independent inputs of quantities computed from the guide and from values of the potential, so it can be estimated and optimized without evaluating the probabilities of the law induced by the guide.
What remains is the size of the search space: a guide is a $[0,1]^d$-valued function of $d$ continuous inputs.

\paragraph{Low-degree approximation and coarse estimation (\cref{sec:estimatorconstruction}).}
Such a function can depend on exponentially many interactions among its inputs.
We restrict the search to functions whose coordinates have \emph{interaction degree} at most $D$: every nonzero component of their Hoeffding decompositions~\cite[\S8.3]{ODo14} involves at most $D$ inputs (\cref{sec:hoeffding}).
Once each input is discretized to $\poly(d)$ bins, such a function has $d^{O(D+1)}$ coefficients.
For related uses of low-degree approximation, see~\cite{LMN93,MY26}.
To show that this restricted search contains a good candidate, the argument follows $f\to g\to u$: \cref{sec:guides} turns the exact guide $f$ into a \emph{witness} $g$, a guide with a small objective gap $\log Z_\Phi-\cP(g)\le b_*$ and small degree-$D$ tails~\eqref{alg:guidehyp}, and \cref{sec:estimatorconstruction} discretizes and truncates $g$ into a candidate $u$ that is nearly as good.
Crucially, the algorithm never evaluates $f$ or $g$ explicitly; their existence merely certifies that the low-degree search space contains a near-optimal candidate $u$.
Instead, the algorithm optimizes the low-degree Walsh coefficients of $u$ directly via projected stochastic supergradient ascent~\cite{NJLS09}.
A penalty for values outside $[0,1]$ keeps the objective at most $\log Z_\Phi$ for every candidate.
We estimate this value on fresh samples and subtract an error margin, giving a lower estimate with high probability.
The additive error is $b_*+O(1)$ at cost $\exp(O((D+1)\log d))$ up to polynomial factors (\cref{alg:generic}).
It remains to construct witnesses.

\paragraph{Ordinary resampling and the square-root baseline (\cref{sec:resampling}).}
The exact guide has objective gap zero, but its moving thresholds can create interactions of every order, so it need not be a witness.
Ordinary resampling replaces each input independently with probability $\tau$, which multiplies a component involving $k$ inputs by $(1-\tau)^k$~\cite[\S8.3]{ODo14} and costs at most $\tau\Hs(\mu)\le d\tau\log2$ of objective value (\cref{guide:resampling}).
Taking $D=O(\tau^{-1}\log(d+B))$ gives exponential cost $\exp(O(d\tau+Q/\tau))$, where $Q=\log d\,\log(d+B)$, and balancing the terms gives the square-root baseline of \cref{alg:baseline}: $\exp(O(\sqrt n\log n))$ time for \BIS\ at fixed accuracy, which already answers the question of~\cite{JPP23}.
The loss is linear in $\tau$ because a resampled input contributes nothing to the linear score.

\paragraph{Threshold smoothing and the cube-root bound (\cref{sec:smoothing}).}
Threshold smoothing keeps every input and shifts its threshold instead.
For a smoothing parameter $0<w\le1/2$, each threshold $c$ encountered by the exact guide becomes $c\pm wc(1-c)$ with a fresh fair sign.
The perturbed decisions determine the next threshold.
Averaging the signs at fixed inputs, with one sign sequence shared by all output coordinates, gives the guide $g$ of~\eqref{sm:perturbedencoder}.
The two shifted thresholds have mean $c$, which preserves each branching probability and hence the joint law $\mu$.
Concavity therefore protects the potential term.
Symmetry cancels the first-order score change, while the factor $c(1-c)$ compensates for the entropy curvature $-\Hs''(c)=1/[c(1-c)]$.
The linear-score loss is at most $w^2c(1-c)$ per comparison and $dw^2/4$ in total (\cref{sm:encoder}).
For a fixed threshold $c$, the averaged indicator has variance $(1-w/2)c(1-c)$ (\cref{sm:kernel}), a factor $1-w/2$ smaller than the original indicator's variance.
An exact energy identity for threshold trees~\eqref{sm:identity} turns this into geometric decay of the high-degree tails at a rate linear in $w$.
This is an aggregate statement rather than a componentwise contraction.
Thus $D=O(w^{-1}\log(d+B))$ suffices, the distortion costs the walk $\exp(O(dw^2+1))$, and each estimate costs $\exp(O(Q/w))$.
Choosing $w$ of order $(Q/d)^{1/3}$ when this is at most $1/2$, and using exact enumeration otherwise, gives the exponent $O(\min\{d,d^{1/3}Q^{2/3}\})$ of \cref{thm:lsm}.

\subsection{Related work}\label{sec:related}

\paragraph{Algorithms for independent sets.}
Ge and \v{S}tefankovi\v{c}~\cite{GS12} proposed a Markov chain for \BIS\ on edge subsets, with weights determined by adjacency-matrix ranks over $\mathbb F_2$.
Goldberg and Jerrum~\cite{GJ12GS} later proved exponential mixing time on an infinite family of bipartite graphs.

For fixed $\Delta\ge3$ and fixed $0<\lambda<\lambda_c(\Delta)$, a degree bound $\Delta$ on one side suffices for nearly linear field-dynamics sampling and nearly cubic Glauber mixing~\cite{CLY23}.
Faster Glauber mixing from a fully occupied side is known with degree balance~\cite{CF25} or with no degree bound on the other side~\cite{FY26}.
At criticality, results require a bound on the maximum degree of the whole graph~\cite{CCYZ25,CJ25} and draw on spectral independence and field dynamics~\cite{ALO20,CFYZ21}.

Low-temperature methods include polymer and clique dynamics~\cite{CGGPSV21,FGKP23}, perfect sampling in an unbalanced regime~\cite{BCP24}, and improved fugacity bounds on regular bipartite expanders~\cite{JMP26}.
For sufficiently large fixed $\Delta$, algorithms cover fugacities down to $\tilde\Omega(1/\Delta)$, including $\lambda=1$, on almost every $\Delta$-regular bipartite graph~\cite{JKP20,LLLM22,CGSV22}.

For every fixed $c>0$, the randomized container algorithm of~\cite[Theorem~3]{JPP23} achieves $n^{-c}$ relative error with probability at least $2/3$ in time $\exp(O(n\log^3\Delta/\Delta))$ on $\Delta$-regular bipartite graphs.
Its guarantee does not cover arbitrary accuracy~\cite{CDKP24}.
Pinning can also destroy regularity, obstructing the usual sampling self-reduction~\cite{JPP23}.
Our estimator handles every pinned residual, yielding counting and sampling at arbitrary accuracy (\cref{conv:conversion}).

\paragraph{Cardinality-constrained independent sets.}
Counting bipartite independent sets of a given total size or occupation of one side is NP-hard to approximate within any polynomial factor when the prescription is part of the input~\cite{CDFGL19}.
Further hardness results hold at certain fixed occupation densities on bounded-degree bipartite graphs~\cite{NPWW26}.
Such constraints generally destroy log-supermodularity, even after complementing one side.

\paragraph{Self-reduction and marginal estimation.}
Bounded self-reductions~\cite{CGMW26} and rejection sampling on self-reduction trees~\cite{HL08,HMJ13} require upper bounds on residual masses that remain consistent under recursion.
Moitra~\cite{Moi19} estimates conditional marginals by a coupling-based linear program for bounded-degree CNFs in the local-lemma regime.
We use the Sinclair--Jerrum tree walk~\cite[Section~4]{SJ89} with coarse variational lower estimates of log-partition functions, without requiring consistency between parent and child estimates.
Samples yield relative counts by self-reduction~\cite{JVV86}.
Our estimates apply even when bipartite hard-core models have coexisting phases and exponentially slow local dynamics~\cite{DFJ02,MWW09}.

\paragraph{Variational bounds on partition functions.}
Bethe bounds are lower bounds for Boolean models with log-supermodular factors~\cite{Ruozzi12}; other approaches use convex hierarchies, mean-field approximation, and correlation rounding~\cite{Ris16,JKM18,JKR19}.
For an Ising interaction matrix $J$, with Frobenius norm $\|J\|_F$, the additive guarantees of~\cite[Theorem~12 and Remark~13]{JKM18} have leading scale $\epsilon n\|J\|_F$.
Jain, Koehler, and Risteski~\cite{JKR19} bound the mean-field error by $O((n\|J\|_F)^{2/3})$ and give subexponential algorithms with $o(n)$ additive error when $\|J\|_F^2=o(n)$.
Our estimator gives error $\epsilon d+1/8$ for arbitrary positive log-supermodular weights, with weight sizes entering its running time through $B$ (\cref{alg:additive}).

L-FIELD~\cite{DK14,DK15} produces factorized laws from sub- and supergradient bounds; its log-supermodular upper bound has a concave Lov\'asz-extension formulation~\cite[Lemma~3 and Section~4.1]{DK14}.
The logistic perturb-and-MAP upper bound is at least as tight as the L-FIELD upper bound~\cite[Proposition~1]{SB16}.

\paragraph{Sequential variational representations.}
For binary variables, \cref{lat:descent} is equivalent to Hazan and Jaakkola's random-perturbation identity~\cite[Theorem~1]{HJ12} (\cref{rem:hj}).
Sequential perturbation samplers use upper bounds~\cite{HMJ13}; lower bounds from perturbing a fixed variable subset generally fall below $\log Z$~\cite[Corollary~1]{HMJ13}, while using all variables gives equality but requires one Gumbel perturbation per configuration.
Variational autoregressive networks~\cite{WWZ19} optimize the objective of \cref{lat:treeprinciple} using network branching rules.
It need not be concave even for log-supermodular laws, and its evaluation uses the candidate law's conditional probabilities.
Our fractional relaxation is concave for log-supermodular laws and evaluable under a fixed product law using only guide and weight evaluations, without evaluating the induced law's probabilities.

\paragraph{Organization.}
\Cref{sec:coarseoracle} reduces sampling and counting to coarse mass estimation; \cref{sec:entropy} develops the variational principle and concave relaxation.
\Cref{sec:estimatorconstruction} builds the estimator from guides constructed in \cref{sec:guides} using resampling and threshold smoothing.
\Cref{sec:latticescope} discusses scope and open questions; \cref{sec:finite} gives quantitative conversion and bit implementation details.

\section{Counting and sampling from coarse estimates}\label{sec:coarseoracle}\label{sec:prelim}

We first explain how coarse lower estimates of residual log-partition functions suffice for sampling and counting.
We then derive \cref{thm:lsm} assuming an estimator with the accuracy and running-time guarantees stated below.
To construct such an estimator, \cref{sec:entropy} develops a concave variational formulation, which \cref{sec:estimatorconstruction} turns into a finite-dimensional optimization problem.
\Cref{sec:guides} uses threshold smoothing to establish the approximation bounds needed for the stated guarantees.

\paragraph{Notation and computational model.}\label{sec:model}\label{sec:prelimmodel}
All logarithms are natural unless a base is specified.
Write $[k]=\{1,\ldots,k\}$, let $e_j$ be the $j$th unit vector, and identify subsets with their indicator vectors.
We use $x_{<i}$ and $x_{\le i}$ for prefixes, and $\ind\{\cdot\}$ and $\ind_A$ for indicators.
For finite laws, $\|\mu-\nu\|_{\TV}=\frac12\sum_x|\mu(x)-\nu(x)|$.
Graphs are finite, simple, and labeled.

We work in an \emph{exact real-arithmetic model}.
Arithmetic operations, comparisons, and evaluations of $\log$ and $\exp$ on real numbers have unit cost, as does an independent uniform draw from $(0,1)$.
We also charge one operation for each accessed coefficient or graph entry, and $T_W$ operations for each evaluation of a weight $W$.
The model has no oracle for partition functions, conditional probabilities or expectations: the algorithm estimates expectations from independent samples and computes every function it queries.
Running-time bounds count operations and random draws on every execution.
\Cref{sec:finite} implements the algorithm with bit operations and fair random bits at the claimed cost.
\subsection{The self-reduction tree}\label{sec:selfreduction}

Let $W:\{0,1\}^d\to\Q_{>0}$ satisfy the evaluator and value-length assumptions of \cref{thm:lsm}, and fix an order of its $d\ge2$ coordinates.
The estimator bounds use this original dimension $d$, including for residuals with fewer than two free coordinates.

The \emph{self-reduction tree} is the complete binary tree of depth $d$ whose nodes $v$ at depth $j$ are the prefixes $x_v\in\{0,1\}^j$; it is a special case of the self-reducibility tree of Sinclair and Jerrum~\cite[p.~100]{SJ89}.
We write $\depth(v)=j$.
The root $o$ is the empty prefix, the leaves are the configurations $x$, $v^-$ denotes the parent of a node $v\ne o$, and $v0$ and $v1$ denote the children of a node $v$ with $\depth(v)<d$.

Pinning the prefix $x_v$ at a node $v$ of depth $j$ leaves the \emph{residual} weight $W_v(x')=W(x_v,x')$ on the free coordinates $x'\in\{0,1\}^{d-j}$.
Its partition function $Z_v=\sum_{x'}W_v(x')$ is the \emph{mass} of $v$, and its \emph{residual law} $W_v/Z_v$ is the conditional law of the free coordinates given the prefix $x_v$.
Thus $Z_o=Z_W$, the residual law of the root is $\mu_W$, and $Z_x=W(x)$ at a leaf $x$.
Pinning the next coordinate to $0$ or $1$ gives the two child residuals, so $Z_v=Z_{v0}+Z_{v1}$ at every internal node.
If $W$ is log-supermodular, so is every residual, since \eqref{wt:condition} remains valid when the prefix coordinates are fixed.
Let $X$ have law $\mu_W$.
The \emph{conditional marginal probability} at a node $v$ with $\depth(v)=i-1<d$ is
\begin{equation}\label{conv:marginal}
p_v=\frac{Z_{v1}}{Z_v} =\mu_W(X_i=1\mid X_{<i}=x_v),
\end{equation}
and we write $p_i(x_{<i})=p_v$ when $x_{<i}=x_v$.
Every prefix has positive probability, so these probabilities are defined at every internal node.
If they were available, we could sample $\mu_W$ by descending the tree, choosing child $v1$ with probability $p_v$.
They also determine the partition function: telescoping the masses along any root-to-leaf path gives the chain rule
\begin{equation}\label{conv:chain}
Z_W=\frac{W(x)} {\prod_{i=1}^d\mu_W(X_i=x_i\mid X_{<i}=x_{<i})} \qquad(x\in\{0,1\}^d).
\end{equation}

For the coarse estimates in the next subsection, we normalize each residual by its value at $0$.
The resulting \emph{normalized potential} is $\Phi_v(x')=\log(W_v(x')/W_v(0))$, with partition function $Z_{\Phi_v}=\sum_{x'}e^{\Phi_v(x')}=Z_v/W_v(0)$.
This rescaling leaves the residual law unchanged.
We have $\Phi_v(0)=0$, the normalization used for the Lov\'asz extension in \cref{sec:lovasz}, and $\log Z_{\Phi_v}\ge0$, because the configuration $x'=0$ contributes $1$ to $Z_{\Phi_v}$.
At a leaf, $Z_{\Phi_v}=1$ and hence $\log Z_{\Phi_v}=0$.

\subsection{From coarse estimates to counting and sampling}
\label{sec:coarseconversion}\label{sec:mainbalance}

We first specify the estimates needed for the counting and sampling conversion.
At a node $v$, the estimator approximates the normalized residual log-partition function $\log Z_{\Phi_v}=\log Z_v-\log W_v(0)$.
Since $W_v(0)$ can be evaluated exactly, adding $\log W_v(0)$ to such an estimate gives an estimate of $\log Z_v$ with the same additive error.

Suppose that, given any node $v$ and any failure parameter $0<\delta_{\rm est}\le1/2$, a randomized \emph{estimator} returns a real $A_v\ge0$ with
\begin{equation}\label{conv:oracle}
\Pr\left[\log Z_{\Phi_v}-b\le A_v\le\log Z_{\Phi_v}\right] \ge1-\delta_{\rm est},
\end{equation}
where $b\ge0$ is a known \emph{one-sided additive error bound} common to all nodes.
Thus, with the stated probability, $A_v$ is a lower bound on $\log Z_{\Phi_v}$ and underestimates it by at most $b$.
The bound $b$ may depend on the original input parameters and the chosen estimator construction; in particular, it may grow with $d$.
At a leaf, $Z_{\Phi_v}=1$, so we require the estimator to return $A_v=0$ exactly.

We also assume a cost bound that is uniform over the nodes.
Every execution, including one on which the accuracy guarantee fails, uses at most $C_0\poly(d,\log(1/\delta_{\rm est}))$ operations and uniform draws, including evaluations of $W$.
The factor $C_0\ge T_W$ absorbs the dependence on the estimator's other construction parameters, which are fixed in advance.
It may depend on the original input parameters and need not be polynomial in $d$, but it is independent of the node and the requested failure probability.
The condition $C_0\ge T_W$ also pays for evaluating the exact leaf weights and the normalizing value $W_v(0)$, as needed by the conversion.

\begin{proposition}[Counting and sampling from coarse estimates]
\label{conv:conversion}
Suppose every residual admits the estimator in~\eqref{conv:oracle}, with additive error bound $b$ and cost $C_0\poly(d,\log(1/\delta_{\rm est}))$.
For $\eta,\epsilon,\delta\in(0,1/2]$, every residual law can be sampled within total variation distance $\eta$ at cost
\[
\exp(O(b+1))\,C_0\poly(d,\log(1/\eta)),
\]
and $Z_W$ can be estimated within relative error $\epsilon$, with failure probability at most $\delta$, at cost
\[
\exp(O(b+1))\,C_0 \poly(d,\epsilon^{-1},\log(1/\delta)).
\]
These bounds hold on every execution.
No log-supermodularity is needed for this conversion.
\end{proposition}

\paragraph{Why coarse estimates suffice.}
Direct self-reduction by descent~\cite{JVV86} would use ratios of mass estimates in place of the conditional probabilities in~\eqref{conv:marginal}.
If $e^{-b}Z_y\le q_y\le Z_y$ for both children $y$ of $v$, then $e^{-b}p_v\le q_{v1}/(q_{v0}+q_{v1})\le e^bp_v$.
These errors multiply along the path, up to a factor $e^{bd}$, so obtaining relative error $\epsilon$ from this product bound calls for $b=O(\epsilon/d)$.
The walk of Sinclair and Jerrum~\cite[Section~4]{SJ89} pays for the error once, in its running time.
The sampling half of \cref{conv:conversion} is a quantitative form of their theorem~\cite[Theorems~4.4 and~4.5]{SJ89}: for self-reducible relations, counts within a polynomial ratio $r$ suffice for almost uniform generation.
They weight each edge of the tree by the estimated number of solutions below it, and run the lazy walk that, when it moves, chooses an incident edge in proportion to its weight.
Its stationary probabilities are proportional to the sums of incident edge weights.
Single solutions are counted exactly, so the stationary law conditioned on the leaves is uniform.
To obtain samples efficiently, it remains to control both the time to approach stationarity and the probability that the endpoint is a leaf.
On a tree of depth $d$, the leaves carry stationary probability at least $1/(2rd)$~\cite[Lemma~4.1]{SJ89}, which controls the number of runs needed to obtain a leaf endpoint.
Each subtree hangs from a single edge whose weight is within the ratio $r$ of the number of leaves below it, so the conductance is at least $1/(4r^2d)$~\cite[Lemma~4.2]{SJ89}.
Their bound on mixing in terms of conductance then controls the length of each run~\cite[Theorem~3.4 and Lemma~4.3]{SJ89}.

\paragraph{What changes here.}
We need this conversion for weighted leaves and for approximation ratios $r=e^b$ that may be superpolynomial in $d$.
To turn our logarithmic estimates into edge weights, we use the mass estimates $q_v=W_v(0)e^{A_v}$, which satisfy $q_x=W(x)$ at leaves and $e^{-b}Z_v\le q_v\le Z_v$ when all estimates are good.
The arguments for the stationary law, leaf mass and conductance bound above use only these two properties, so they carry over with leaf weights $W(x)$ and $r=e^b$.
In particular, the stationary law conditioned on the leaves is $\mu_W$.
For the running time, however, their bound through the minimum stationary probability could depend on very small leaf weights.
We avoid this dependence by bounding convergence from the root using its stationary probability, which is at least $1/(2r^2d)$.
This gives mixing time polynomial in $r$, $d$, and $\log(1/\eta)$.
Together with the leaf-mass bound, it gives an $e^{O(b)}$ sampling overhead up to polynomial factors in $d$ and $\log(1/\eta)$, even when $b$ grows with $d$.
Randomized estimates are computed once and cached, as in~\cite[Remark after Theorem~4.5]{SJ89}, so that the analysis can condition on fixed edge weights.
Since the same construction applies to every residual, the resulting samples give a relative counter by self-reduction~\cite{JVV86}.
\Cref{sec:conversiondetails} gives the full construction and failure-probability analysis.

\paragraph{The log-supermodular estimator.}
The following estimator supplies the estimates required by \cref{conv:conversion}; its proof is completed in \cref{sec:smoothing}.

\begin{theorem}[Estimator from threshold smoothing]\label{alg:coarsetheorem}
Assume $W$ is log-supermodular and let $0<w\le1/2$.
For every node $v$ and every real $0<\delta\le1/2$, a randomized algorithm returns a real $A_v\in[0,d(1+2B)]$ such that
\begin{equation}\label{alg:coarseguarantee}
\Pr\Bigl[\log Z_{\Phi_v}-\Bigl(\frac{dw^2}4+\frac18\Bigr)\le A_v\le\log Z_{\Phi_v}\Bigr] \ge1-\delta .
\end{equation}
Every execution uses at most
\begin{equation}\label{alg:coarsecost}
\exp\bigl(O(w^{-1}\log d\,\log(d+B))\bigr)\poly\bigl(d,B,T_W,\log(1/\delta)\bigr)
\end{equation}
operations and uniform draws.
At leaves it returns $A_v=0$.
\end{theorem}
The estimator subtracts an error margin from an estimated objective value; the objective itself never exceeds $\log Z_{\Phi_v}$.

\paragraph{The cube-root balance.}
The parameter $w$ trades accuracy against cost.
In the additive error bound $b=dw^2/4+1/8$, the term $dw^2/4$ bounds how far the value of the guide constructed in \cref{sec:guides} falls below $\log Z_{\Phi_v}$, and $1/8$ bounds the further loss of discretization, truncation, optimization and validation in \cref{sec:estimatorconstruction}.
The error bound $b$ contributes $\exp(O(dw^2+1))$ to the walk, whereas each estimate costs $\exp(O(Q/w))$ up to polynomial factors, with $Q=\log d\,\log(d+B)$.
Balancing $dw^2$ and $Q/w$ yields $d^{1/3}Q^{2/3}$.

\begin{proof}[Proof of \cref{thm:lsm}]
Put $\ell_d=\ceil{\log_2 d}$ and $\ell_B=\ceil{\log_2(d+B)}$.
Suppose first that $\ell_d\ell_B<d/8$, and let $w$ be the largest number $2^{-j}$, $j\ge1$, with $dw^3\le\ell_d\ell_B$; it is found by comparing $d$ with $\ell_d\ell_B8^j$ for $j=1,2,\ldots$.
Then $w<1/2$, and maximality gives $8dw^3>\ell_d\ell_B$, whence
\[
dw^2\le d^{1/3}(\ell_d\ell_B)^{2/3},\qquad w^{-1}\ell_d\ell_B<2d^{1/3}(\ell_d\ell_B)^{2/3}.
\]
By \cref{alg:coarsetheorem}, the estimator with this parameter satisfies~\eqref{conv:oracle} with $b=dw^2/4+1/8$ and $C_0=\exp(O(w^{-1}\ell_d\ell_B))\poly(d,B,T_W)$.
\Cref{conv:conversion} then gives the required sampler and relative counter, with exponential factor $\exp(O(d^{1/3}(\ell_d\ell_B)^{2/3}))$.

If instead $\ell_d\ell_B\ge d/8$, then $d\le4d^{1/3}(\ell_d\ell_B)^{2/3}$, and we evaluate all $2^d$ weights, sum them, and sample by comparing one uniform draw with their cumulative sums.
This exact enumeration costs $2^d\poly(d,T_W)$.
In both cases the exponent is $O(\min\{d,d^{1/3}(\ell_d\ell_B)^{2/3}\})$.
Since $\ell_d\le2\log_2 d$ and $\ell_B\le2\log_2(d+B)$, this gives the bound of \cref{thm:lsm}.
\end{proof}

At the root, the same estimator gives additive estimates directly.

\begin{corollary}[Additive estimates]\label{alg:additive}
Let $W$ be log-supermodular.
For all reals $0<\epsilon\le1/16$ and $0<\delta\le1/2$, a randomized algorithm returns a real $A$ with $\Pr[\log Z_W-\epsilon d-1/8\le A\le\log Z_W]\ge1-\delta$.
Every execution uses at most $\exp(O(\epsilon^{-1/2}\log d\,\log(d+B)))\poly(d,B,T_W,\log(1/\delta))$ operations and uniform draws.
\end{corollary}
\begin{proof}
Apply \cref{alg:coarsetheorem} at the root with $w=2\sqrt\epsilon\le1/2$, so that $dw^2/4=\epsilon d$, and add $\log W(0)$, since $Z_{\Phi_o}=Z_W/W(0)$.
\end{proof}

Such an additive estimate at the root is not a relative approximation of $Z_W$.
By \cref{conv:conversion}, lower estimates within additive error polylogarithmic in $d+B$ at every node, at quasi-polynomial cost, would already give a randomized quasi-polynomial approximation scheme.

The next three sections prove \cref{alg:coarsetheorem}.
\Cref{sec:entropy} writes $\log Z_{\Phi_v}$ as the maximum of an objective that is concave for log-supermodular weights, \cref{sec:estimatorconstruction} maximizes a finite-dimensional version of it from samples, and \cref{sec:guides} shows that this version loses little.

\section{A variational principle on the self-reduction tree}\label{sec:entropy}

We develop a variational formulation for constructing the coarse lower estimates of the residual log-partition functions $\log Z_{\Phi_v}$ required by~\eqref{conv:oracle}.
Starting from Gibbs' variational principle on the self-reduction tree (\cref{sec:treeprinciple}), we linearize the entropy terms (\cref{sec:descents}) and use the Lov\'asz extension for the potential term (\cref{sec:lovasz}).
The resulting objective has maximum $\log Z_{\Phi_v}$, so every feasible choice gives a lower bound; it is concave when the weight is log-supermodular.

We state the formulation for a general potential $\Phi:\{0,1\}^d\to\R$ normalized by $\Phi(0)=0$.
Write $Z_\Phi=\sum_xe^{\Phi(x)}$ and $\mu=e^\Phi/Z_\Phi$.
Every strictly positive law on the cube has this form for a unique normalized potential, namely $\Phi=\log(\mu/\mu(0))$.
The tree masses and conditional marginal probabilities in this section are those of the weight $e^\Phi$.

\subsection{Branching rules}\label{sec:treeprinciple}

Sampling by self-reduction descends the tree from the root, moving from each internal node $v$ to $v1$ with probability $p_v$ and to $v0$ otherwise.
To obtain $p$ as the solution of an optimization problem, we let these probabilities vary.
A \emph{branching rule} $\theta$ assigns a probability $\theta_v\in[0,1]$ to every internal node $v$ of the self-reduction tree, that is, to every node with $\depth(v)<d$.
We write $\theta_x$ for the branching probability at the node with prefix $x$, so that $\theta_o$ is the branching probability at the root~$o$.

\begin{figure}[ht]
\centering
\begin{tikzpicture}[ level distance=12mm, level 1/.style={sibling distance=60mm}, level 2/.style={sibling distance=30mm}, every node/.style={font=\small}, lab/.style={font=\footnotesize,inner sep=1.5pt}]
\node {$o$} child {node {$0$} child {node[label={[font=\footnotesize]below:$(1-\theta_o)(1-\theta_0)$}] {$00$} edge from parent node[lab,above left] {$1-\theta_0$}} child {node[label={[font=\footnotesize]below:$(1-\theta_o)\,\theta_0$}] {$01$} edge from parent node[lab,above right] {$\theta_0$}} edge from parent node[lab,above left] {$1-\theta_o$}} child {node {$1$} child {node[label={[font=\footnotesize]below:$\theta_o(1-\theta_1)$}] {$10$} edge from parent node[lab,above left] {$1-\theta_1$}} child {node[label={[font=\footnotesize]below:$\theta_o\,\theta_1$}] {$11$} edge from parent node[lab,above right] {$\theta_1$}} edge from parent node[lab,above right] {$\theta_o$}};
\end{tikzpicture}
\caption{A branching rule on the self-reduction tree for $d=2$.
Nodes other than the root $o$ are labeled by their prefixes, and edges by the probabilities with which the descent takes them.
The probability of reaching each leaf, shown below it, is the product along its path, and by~\eqref{lat:entropychain} the entropy of the leaf law is $\Hs(\theta_o)+(1-\theta_o)\Hs(\theta_0)+\theta_o\Hs(\theta_1)$.}
\label{fig:branching}
\end{figure}
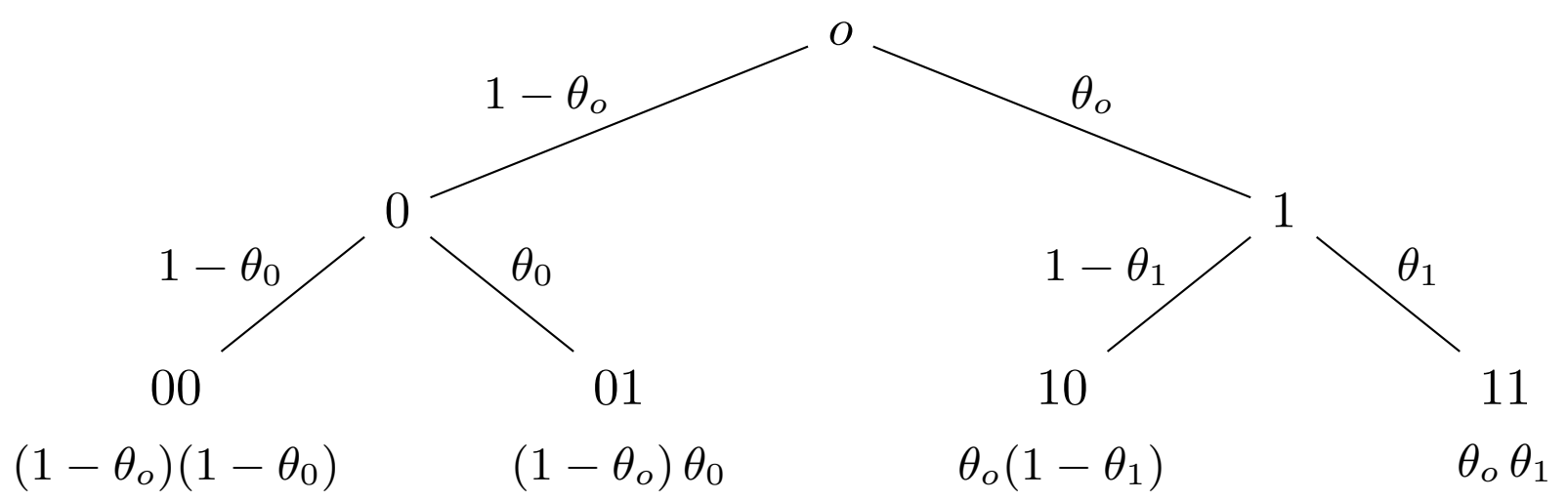

The \emph{descent} with rule $\theta$ starts at the root and moves from each internal node $v$ to $v1$ with probability $\theta_v$ and to $v0$ otherwise (\cref{fig:branching}).
It reaches the node with prefix $x\in\{0,1\}^j$ with probability
\begin{equation}\label{lat:reach}
\nu_\theta(x)=\prod_{i\le j:\,x_i=1}\theta_{x_{<i}}\prod_{i\le j:\,x_i=0}\bigl(1-\theta_{x_{<i}}\bigr),
\end{equation}
the product of the branching probabilities along the path.
We write $\nu_\theta(v)$ for the probability of reaching node $v$ and $\nu_\theta$ for the induced law on leaves.
Every law $\nu$ on $\{0,1\}^d$ is $\nu_\theta$ for some $\theta$: at each node that $\nu$ reaches, let $\theta_v$ be the conditional probability under $\nu$ that the next coordinate is $1$, and choose arbitrary values elsewhere.
In particular $\nu_p=\mu$ by~\eqref{conv:chain}.

To apply Gibbs' variational principle, we express the entropy of $\nu_\theta$ in terms of the branching probabilities.
The \emph{entropy} of a law $\nu$ on a finite set is $\Hs(\nu)=-\sum_x\nu(x)\log\nu(x)$, and the \emph{binary entropy} of $p\in[0,1]$ is $\Hs(p)=-p\log p-(1-p)\log(1-p)$, where $0\log0=0$~\cite[Chapter~2]{CT06}.
Let $X$ have law $\nu_\theta$.
Given $X_{<i}$, the coordinate $X_i$ is a Bernoulli variable with mean $\theta_{X_{<i}}$, so the chain rule for entropy~\cite[Theorem~2.5.1]{CT06} gives
\begin{equation}\label{lat:entropychain}
\Hs(\nu_\theta)=\sum_{i=1}^d\Hs(X_i\mid X_{<i}) =\sum_{i=1}^d\E\,\Hs\bigl(\theta_{X_{<i}}\bigr) =\sum_v\nu_\theta(v)\Hs(\theta_v),
\end{equation}
the last sum over internal nodes.
At each depth, the probabilities $\nu_\theta(v)$ sum to one, so that depth contributes the expected binary entropy of the next branch.

For every law $\nu$ on the cube, Gibbs' variational principle, in the form of~\cite[Theorem~3.4 and \S5.2.2]{WJ08}, gives
\begin{equation}\label{pre:gibbs}
\Hs(\nu)+\E_\nu\Phi=\log Z_\Phi-\mathrm{KL}(\nu\Vert\mu)\le\log Z_\Phi, \qquad \mathrm{KL}(\nu\Vert\mu)=\sum_x\nu(x)\log\frac{\nu(x)}{\mu(x)}\ge0,
\end{equation}
with equality exactly when $\nu=\mu$; here $\mathrm{KL}(\nu\Vert\mu)$ is the \emph{relative entropy} of $\nu$ with respect to $\mu$, and terms with $\nu(x)=0$ are zero.
Substituting~\eqref{lat:entropychain} for the entropy of $\nu=\nu_\theta$ writes the principle on the tree.

\begin{proposition}[Variational principle on the self-reduction tree]\label{lat:treeprinciple}
For every strictly positive law $\mu=e^\Phi/Z_\Phi$ on $\{0,1\}^d$,
\begin{equation}\label{lat:branching}
\log Z_\Phi=\max_\theta\Bigl(\sum_v\nu_\theta(v)\Hs(\theta_v)+\E_{\nu_\theta}\Phi\Bigr),
\end{equation}
the maximum over branching rules $\theta$, and the unique maximizer is the vector $\theta=p$ of conditional marginal probabilities of $\mu$.
\end{proposition}
\begin{proof}
By~\eqref{lat:entropychain} and~\eqref{pre:gibbs}, the bracket equals $\Hs(\nu_\theta)+\E_{\nu_\theta}\Phi=\log Z_\Phi-\mathrm{KL}(\nu_\theta\Vert\mu)$, which is maximal exactly when $\nu_\theta=\mu$.
Since $\mu$ is strictly positive, every internal node $v$ is reached with positive probability, so $\theta_v=\nu_\theta(X_i=1\mid X_{<i}=x_v)=p_v$ for $i-1=\depth(v)$.
\end{proof}

The maximizer is the vector of conditional marginal probabilities that self-reduction needs, and by~\eqref{pre:gibbs} every branching rule gives a lower bound.
This formulation does not yet yield the desired algorithm.
It has $2^d-1$ variables, and the probabilities $\nu_\theta(v)$ in~\eqref{lat:reach} are products along paths.
The objective is not concave in $\theta$, even for log-supermodular laws: for $d=2$ and $\Phi(x)=3x_1x_2$, its Hessian in the two branching probabilities on the path to $(1,1)$ has determinant $-1$ at $\theta\equiv\frac12$.

Using the probabilities $\nu(v)$, or equivalently the leaf law $\nu$, as variables restores concavity, but leaves an evaluation problem.
Direct Monte Carlo evaluation of $\Hs(\nu)=\E_\nu[-\log\nu(X)]$ requires sampling from the candidate law and evaluating its probabilities.
Rewriting this expectation under the uniform law on leaves introduces the weights $2^d\nu(x)$, and the resulting integrand can have exponentially large second moment.
We next drive every descent by the same independent uniform inputs.
Thresholding a fresh input at each step expresses the entropy as an expectation under this fixed law.
The expectation is linear in the map computing the descent and requires no evaluation of the induced leaf probabilities.

\subsection{Linearizing the entropy}\label{sec:descents}

The linearization rests on an identity for a single bit.
Let $U$ be uniform on $(0,1)$, and let $a(u)=\log(u/(1-u))$ be the \emph{logit function}.
It is integrable, strictly increasing and antisymmetric, $a(1-u)=-a(u)$, so $\E a(U)=0$.
Since $\Hs'=-a$ on $(0,1)$,
\begin{equation}\label{lat:bernoulli}
\int_c^1a(u)\,du=\Hs(c)\qquad(0\le c\le1).
\end{equation}
We call $\E[a(U)\phi(U)]=\int_0^1a\phi$ the \emph{linear score} of a bounded measurable function $\phi$ on $(0,1)$.
The \emph{threshold indicator} $\ind_{[c,1)}$, where $c=1-p$, equals $1$ at $U$ with probability $p$, and by~\eqref{lat:bernoulli} its linear score is the entropy of that bit:
\begin{equation}\label{lat:onebit}
\E\bigl[a(U)\ind\{U\ge1-p\}\bigr]=\Hs(p)\qquad(0\le p\le1).
\end{equation}
The left side is linear in the indicator, whereas $\Hs(p)$ is concave in $p$.
No other $[0,1]$-valued function with the same mean scores as much; this is the bathtub principle~\cite[Theorem~1.14]{LL01} with its equality case.

\begin{lemma}[Bathtub principle]\label{lat:rearrangement}
If $\phi:(0,1)\to[0,1]$ is measurable and $\int_0^1\phi=p$, then $\int_0^1a\phi\le\Hs(p)$, with equality only if $\phi=\ind_{[1-p,1)}$ almost everywhere.
\end{lemma}
\begin{proof}
For $p\in\{0,1\}$, $\phi$ equals the constant $p=\ind_{[1-p,1)}$ almost everywhere, and $\int_0^1a\phi=0=\Hs(p)$.
Otherwise put $c=1-p$ and $\psi=\ind_{[c,1)}-\phi$, so that $\int\psi=0$, $\psi\ge0$ on $[c,1)$ and $\psi\le0$ on $(0,c)$.
Since $a$ is increasing,
\[
\Hs(p)-\int_0^1a\phi=\int_0^1a\psi=\int_0^1\bigl(a(u)-a(c)\bigr)\psi(u)\,du\ge0 .
\]
The last integrand is nonnegative, and since $a$ is strictly increasing it vanishes only where $u=c$ or $\psi(u)=0$; so equality forces $\psi=0$ almost everywhere.
\end{proof}

The identity~\eqref{lat:onebit} is also the one-bit case of Hazan and Jaakkola's random-perturbation identity (\cref{rem:hj}).

\paragraph{Encoding a descent.}
To apply~\eqref{lat:onebit} at every step of a descent, let $U_1,\ldots,U_d$ be independent uniform inputs on $(0,1)$, one for each depth, and write $U=(U_1,\ldots,U_d)$.
The \emph{threshold descent} $f^\theta$ of a branching rule $\theta$ compares each fresh input with $1-\theta_v$ at the current node $v$.
Let $f=f^p$ be the threshold descent using the conditional marginal probabilities of $\mu$:
\begin{equation}\label{lat:encoder}
f^\theta_i(U)=\ind\bigl\{U_i\ge1-\theta_{f^\theta_{<i}(U)}\bigr\},\qquad f_i(U)=f^p_i(U)=\ind\bigl\{U_i\ge1-p_i(f_{<i}(U))\bigr\}.
\end{equation}
Because $U_i$ is uniform on $(0,1)$ and independent of $U_{<i}$, the conditional probability that $f^\theta_i(U)=1$ given $U_{<i}$ is the branching probability of the node $f^\theta_{<i}(U)$, so $f^\theta(U)$ has law $\nu_\theta$; in particular $f(U)$ has law $\mu$.
The map $f$ performs sequential inverse-transform sampling, the discrete analogue of the inverse Rosenblatt transformation~\cite{Ros52}.

\begin{lemma}[Entropy identity for threshold descents]\label{lat:exactencoder}
For every branching rule $\theta$, the threshold descent $f^\theta$ takes values in $\{0,1\}^d$, its $i$th coordinate depends only on $U_1,\ldots,U_i$, and $\E[a(U_i)f^\theta_i(U)\mid U_{<i}]=\Hs(\theta_{f^\theta_{<i}(U)})$ for every $i$.
Hence $\sum_i\E[a(U_i)f^\theta_i(U)]=\Hs(\nu_\theta)$, and adding $\E\Phi(f^\theta(U))$ gives the bracket in~\eqref{lat:branching}.
In particular $\sum_i\E[a(U_i)f_i(U)]=\Hs(\mu)$, and
\[
\sum_{i=1}^d\E[a(U_i)f_i(U)]+\E\Phi(f(U))=\log Z_\Phi .
\]
\end{lemma}
\begin{proof}
The first two claims are immediate from~\eqref{lat:encoder}.
The identity is~\eqref{lat:onebit} with $p=\theta_{f^\theta_{<i}(U)}$, which is a function of $U_{<i}$.
Since $f^\theta(U)$ has law $\nu_\theta$, summing over $i$ gives $\sum_v\nu_\theta(v)\Hs(\theta_v)=\Hs(\nu_\theta)$ by~\eqref{lat:entropychain}, and $\E\Phi(f^\theta(U))=\E_{\nu_\theta}\Phi$.
For $\theta=p$, \cref{lat:treeprinciple} gives the value $\log Z_\Phi$.
\end{proof}

\begin{example}[Threshold descent on a path]\label{ex:path}
Let $G$ be the path $r_1\,v_1\,r_2\,v_2\,r_3$ with $L=\{v_1,v_2\}$, in this order, and $R=\{r_1,r_2,r_3\}$.
Summing out $R$ gives $W(\varnothing)=8$, $W(\{v_1\})=W(\{v_2\})=2$, and $W(\{v_1,v_2\})=1$, so $Z(G)=13$.
The conditional marginal probabilities are $p_1=3/13$, $p_2(0)=1/5$, and $p_2(1)=1/3$.
Thus $f$ is given by
\[
f_1(U)=\ind\{U_1\ge10/13\},\qquad f_2(U)=\begin{cases}
\ind\{U_2\ge4/5\},&U_1<10/13,\\
\ind\{U_2\ge2/3\},&U_1\ge10/13.
\end{cases}
\]
By \cref{lat:exactencoder}, with $\Phi=\log(W/8)$, $\sum_i\E[a(U_i)f_i(U)]+\E\Phi(f(U))=\log Z_\Phi=\log(13/8)$.
The second coordinate illustrates an interaction: its threshold for $U_2$ moves with the decision on $U_1$, a dependence measured in \cref{sec:hoeffding} and reduced by averaging in \cref{sec:guides}.
\end{example}

\paragraph{From threshold descents to guides.}
By \cref{lat:exactencoder}, the bracket in~\eqref{lat:branching} is a score that is linear in the descent $f^\theta$, plus an expected potential.
This represents the objective under a fixed product law without evaluating the induced leaf probabilities.
We now optimize over the map computing the descent instead of the branching probabilities.
Threshold descents do not form a convex set: the average of $\ind\{U_1\ge\frac12\}$ and $\ind\{U_1\ge\frac34\}$ equals $\frac12$ on $[\frac12,\frac34)$.
To obtain a convex feasible set we allow maps with values in $[0,1]^d$.

\begin{definition}[Guides]\label{lat:guide}
A \emph{guide} is a measurable map $F:(0,1)^d\to[0,1]^d$ whose $i$th coordinate depends only on $U_1,\ldots,U_i$.
Its coordinates $F_i$ are its \emph{rows}, and its \emph{linear score} is $\sum_{i=1}^d\E[a(U_i)F_i(U)]$.
A guide is \emph{Boolean} if its values lie in $\{0,1\}^d$.
\end{definition}

We write $F$ for a general guide and call $f=f^p$ the \emph{exact guide} of $\mu$.
Every threshold descent is a Boolean guide, and by \cref{lat:exactencoder} its linear score is the entropy of its law.
A general Boolean guide may depend on the full earlier input history and may use an arbitrary measurable set of values of the fresh input to decide its next bit.

We first prove that the variational bound remains valid for all Boolean guides.
This is the form needed in \cref{sec:lovasz}, where general guides are expressed as averages of Boolean guides that need not be threshold descents.
The restriction to earlier and current inputs is essential: choosing $x$ after all inputs are revealed attains $\E\max_x[\Phi(x)+\sum_i a(U_i)x_i]$, which can exceed $\log Z_\Phi$ (\cref{rem:hj}).

\begin{lemma}[Variational principle for Boolean guides]\label{lat:descent}
Let $\mu=e^\Phi/Z_\Phi$ be a strictly positive law on $\{0,1\}^d$.
Every Boolean guide $F$ satisfies
\begin{equation}\label{lat:treebound}
\sum_{i=1}^d\E[a(U_i)F_i(U)]+\E\Phi(F(U))\le\log Z_\Phi,
\end{equation}
with equality if and only if $F=f$ almost everywhere.
Thus $\log Z_\Phi$ is the maximum of the left-hand side over Boolean guides, and the unique maximizer is the exact guide.
\end{lemma}
\begin{proof}
We express the deficit as a sum of two nonnegative losses at each step.
By \cref{sec:selfreduction}, applied to the weight $e^\Phi$, the masses satisfy $Z_o=Z_\Phi$, $Z_v=Z_{v0}+Z_{v1}$ at internal nodes and $\log Z_x=\Phi(x)$ at a leaf $x$, and $p_v=Z_{v1}/Z_v\in(0,1)$.
Write $\mathrm{Ber}(q)$ for the Bernoulli law of mean $q$.
Since $Z_{v0}=(1-p_v)Z_v$ and $Z_{v1}=p_vZ_v$, expanding relative entropy gives the one-step identity
\begin{equation}\label{lat:loss2}
\Hs(q)+(1-q)\log Z_{v0}+q\log Z_{v1} =\log Z_v-\mathrm{KL}\bigl(\mathrm{Ber}(q)\Vert\mathrm{Ber}(p_v)\bigr) \qquad(0\le q\le1).
\end{equation}
The endpoint cases use $0\log0=0$.

Now write $X=F(U)$ and $q_i=\Pr(X_i=1\mid U_{<i})$.
Given $U_{<i}$, the reached node $X_{<i}$ is fixed.
As a function of the fresh input $U_i$, the next bit $X_i$ is the indicator of a measurable set, with mean $q_i$.
Its \emph{rearrangement loss} is nonnegative by \cref{lat:rearrangement}:
\begin{equation}\label{lat:loss1}
R_i=\Hs(q_i)-\E[a(U_i)X_i\mid U_{<i}]\ge0.
\end{equation}
By~\eqref{lat:loss2} at $v=X_{<i}$ with $q=q_i$, and by~\eqref{lat:loss1},
\[
\log Z_{X_{<i}}-\E[\log Z_{X_{\le i}}\mid U_{<i}]-\E[a(U_i)X_i\mid U_{<i}] =R_i+\mathrm{KL}\bigl(\mathrm{Ber}(q_i)\Vert\mathrm{Ber}(p_i(X_{<i}))\bigr).
\]
Taking expectations and summing over $i$ telescopes the masses from $\log Z_o=\log Z_\Phi$ to $\log Z_X=\Phi(X)$, giving
\begin{equation}\label{lat:descentdeficit}
\log Z_\Phi-\sum_{i=1}^d\E[a(U_i)X_i]-\E\Phi(X) =\sum_{i=1}^d\E\left[R_i+ \mathrm{KL}\bigl(\mathrm{Ber}(q_i)\Vert \mathrm{Ber}(p_i(X_{<i}))\bigr)\right].
\end{equation}
Both losses are nonnegative.
They vanish for the exact guide, since its conditional mean is $p_i(f_{<i}(U))$ and each step is the maximizing threshold in \cref{lat:rearrangement}.
Conversely, equality forces $q_i=p_i(X_{<i})$ and $R_i=0$ almost surely for every $i$.
The equality case of \cref{lat:rearrangement} then gives $F_i(U)=\ind\{U_i\ge1-p_i(F_{<i}(U))\}$ almost surely.
Induction using \eqref{lat:encoder} yields $F=f$ almost everywhere.
\end{proof}

Two Boolean guides with the same law can have different linear scores.
The guide $\check f$ with $\check f_i(U)=\ind\{U_i<p_i(\check f_{<i}(U))\}$ samples $\mu$ like the exact guide, but in~\eqref{lat:descentdeficit} its Bernoulli relative entropies vanish and its rearrangement losses total $2\Hs(\mu)$, so its linear score is $-\Hs(\mu)$: the left side of~\eqref{lat:treebound} depends on the guide, not only on its law.

\begin{remark}[Relation to random perturbations]\label{rem:hj}
The identity~\eqref{lat:onebit} is the one-bit case of the random-perturbation identity for log-partition functions.
The logit $a(U)$ has the logistic law~\cite[Section~2]{BS07}, that of a difference of two independent Gumbel variables with mean zero and scale one.
For $0<p<1$ and the log-odds $t=\log(p/(1-p))$, the threshold indicator is the maximizer $\ind\{t+a(U)\ge0\}$ of $x(t+a(U))$ over $x\in\{0,1\}$, so by~\eqref{lat:onebit}
\[
\E\max_{x\in\{0,1\}}x\bigl(t+a(U)\bigr)=pt+\Hs(p)=\log(1+e^t).
\]

For each variable $i$, take independent copies $\gamma_i(0),\gamma_i(1)$ of these Gumbel variables, as in~\cite{HJ12}.
Write $\gamma_i(y_i)=\gamma_i(0)+y_i(\gamma_i(1)-\gamma_i(0))$ for $y_i\in\{0,1\}$.
The additive term $\gamma_i(0)$ vanishes in expectation, leaving the logistic coefficient $\gamma_i(1)-\gamma_i(0)$.
If each variable is chosen after its own coefficient and the earlier ones are revealed, expectations and maxima nest.
Thus the maximization identity in \cref{lat:descent}, for binary variables, is equivalent to Hazan and Jaakkola's random-perturbation identity~\cite[Theorem~1]{HJ12}.
The lemma also identifies the unique maximizer, as needed in \cref{lat:variational}.

If all coefficients are revealed before choosing $x$, the maximum becomes $\E\max_x[\Phi(x)+\sum_i a(U_i)x_i]\ge\log Z_\Phi$, the corresponding upper bound~\cite[Section~4.1, equation~(5)]{HJ12}.
It can be strict: for $d=2$ and $\Phi(x)=3x_1x_2$ it is about $3.187$, whereas $\log Z_\Phi=\log(3+e^3)$ is about $3.139$.
\end{remark}

\subsection{The concave relaxation}\label{sec:lovasz}

Guides form the convex feasible set announced in \cref{sec:descents}.
The linear score is already defined on this set; it remains to extend the potential to fractional points while preserving the variational bound.
By the layer cake representation~\cite[Theorem~1.13]{LL01}, a guide $F$ is the average $\int_0^1\ind\{F>s\}\,ds$ of its \emph{layers} $\ind\{F>s\}=(\ind\{F_i>s\})_{i=1}^d$, which are Boolean guides.
Equivalently, take $U_0$ uniform on $(0,1)$, independently of $U$, and set $X=\ind\{F(U)>U_0\}$.
Then $F_i(U)=\Pr(X_i=1\mid U)$ for every $i$.
We extend the objective in~\eqref{lat:treebound} by averaging over these layers.
For a constant guide $z\in[0,1]^d$, whose linear score is zero, this gives the potential value $\int_0^1\Phi(\ind\{z>s\})\,ds$ at $z$.
This is the Lov\'asz extension~\eqref{L2} below, which is computed from $d+1$ values of $\Phi$ and is concave exactly when $\mu$ is log-supermodular~\eqref{L3}.

We define the extension on all of $\R^d$, because the search of \cref{sec:genericestimator} evaluates it at points outside $[0,1]^d$.
For $z\in\R^d$, order its coordinates as $z_{j_1}\ge\cdots\ge z_{j_d}$, put $S_t=\{j_1,\ldots,j_t\}$, with $S_0=\varnothing$, and define the \emph{Lov\'asz extension}~\cite{Lov83}
\begin{equation}\label{lat:lovasz}
\PL(z)=\sum_{t=1}^dz_{j_t}\bigl(\Phi(S_t)-\Phi(S_{t-1})\bigr);
\end{equation}
ties may be broken arbitrarily.

A \emph{marginal difference} of $\Phi$ is a value $\Phi(x\vee e_j)-\Phi(x)$, and $\Phi$ is \emph{supermodular} if $\Phi(x\wedge y)+\Phi(x\vee y)\ge\Phi(x)+\Phi(y)$ for all $x,y$, that is, if $\mu$ is log-supermodular.
We use five standard facts from~\cite{Lov83} and~\cite[Definition~3.1 and Propositions~3.1, 3.2 and~3.6]{Bach13}, stated here for supermodular functions.
\eqref{L1}, \eqref{L2} and~\eqref{L4} hold for every potential with $\Phi(0)=0$.

\begin{enumerate}[label=(L\arabic*),ref=L\arabic*]
\item\label{L1}
$\PL(\ind_S)=\Phi(S)$ for every set $S$, by telescoping.
\item\label{L2}
For $z\in[0,1]^d$, $\PL(z)=\int_0^1\Phi(\ind\{z>s\})\,ds$, where $\ind\{z>s\}=(\ind\{z_i>s\})_{i=1}^d$: this layer equals $S_t$ for $z_{j_{t+1}}<s<z_{j_t}$ ($0\le t\le d$, with $z_{j_0}=1$ and $z_{j_{d+1}}=0$), and summation by parts gives~\eqref{lat:lovasz}.
\item\label{L3}
If $\Phi$ is supermodular, then $\PL$ is concave on all of $\R^d$, and at every $z$ the vector of chain increments in~\eqref{lat:lovasz}, for any ordering of $z$, is a supergradient.
Conversely, concavity of $\PL$ forces supermodularity.
\item\label{L4}
If every marginal difference of $\Phi$ has absolute value at most $\Lambda$, then so does every chain increment in~\eqref{lat:lovasz}; since $\PL$ is continuous and linear wherever the order of the coordinates is fixed, $|\PL(z)-\PL(z')|\le\Lambda\|z-z'\|_1$.
\item\label{L5}
If $\Phi$ is supermodular and $X$ is a random Boolean vector, then $\E\Phi(X)=\E\PL(X)\le\PL(\E X)$, by~\eqref{L1}, \eqref{L3} and Jensen's inequality.
\end{enumerate}
By~\eqref{L5} and~\eqref{L2}, for supermodular $\Phi$ and $z\in[0,1]^d$, $\PL(z)$ is the largest $\E\Phi(X)$ over random Boolean vectors $X$ with mean $z$.
The maximum is attained by $X=\ind\{z>U_0\}$.
Thus $\PL$ is the concave closure of $\Phi$ on $[0,1]^d$~\cite[Section~5.1]{Bach13}.
For independent sets with the right side summed out, $\Phi(S)=-|N(S)|\log2$ and $\PL(z)=-\log2\sum_{r\in R}\max_{v\in N(r)}z_v$, an empty maximum being zero.

\paragraph{The variational objective.}
Using this extension, define the objective of a guide $F$ by
\begin{equation}\label{lat:score}
\cP(F)=\sum_{i=1}^d\E[a(U_i)F_i(U)]+\E\PL(F(U)).
\end{equation}
The last term is the \emph{potential term}.
We call $\cP$ the \emph{variational objective} and $\cP(F)$ the \emph{objective value} of $F$.
Both terms are finite because $a$ is integrable and $\PL(F(U))$ is bounded.
We call $\log Z_\Phi-\cP(F)$ the \emph{objective gap} of $F$.

\begin{proposition}[Exactness and concavity of the relaxation]\label{lat:variational}
Every guide $F$ satisfies $\cP(F)\le\log Z_\Phi$, with equality if and only if $F=f$ almost everywhere.
The objective is concave on the convex set of these guides if and only if $\Phi$ is supermodular, that is, if and only if $\mu$ is log-supermodular.
\end{proposition}

\begin{proof}Write $F=\int_0^1\ind\{F>s\}\,ds$ as above; for each $s\in[0,1)$ the layer $\ind\{F>s\}$ is a Boolean guide.
The score is linear in $F$, $\PL(F)=\int_0^1\Phi(\ind\{F>s\})\,ds$ by~\eqref{L2}, and by~\eqref{L1} the objective $\cP$ agrees on Boolean guides with the left side of~\eqref{lat:treebound}.
Since $|a(U_i)\ind\{F_i>s\}|\le|a(U_i)|$ is integrable and $\Phi$ is bounded, Fubini's theorem and \cref{lat:descent} give
\begin{equation}\label{lat:layers}
\cP(F)=\int_0^1\cP(\ind\{F>s\})\,ds\le\log Z_\Phi,
\end{equation}
and $\cP(f)=\log Z_\Phi$ by \cref{lat:exactencoder}.
If $\cP(F)=\log Z_\Phi$, then for almost every $s\in[0,1)$ the layer attains equality, so $\ind\{F>s\}=f$ almost everywhere by \cref{lat:descent}.
By Fubini's theorem, for almost every $u$ and every $i$ we have $\ind\{F_i(u)>s\}=f_i(u)$ for almost every $s\in[0,1)$; since $F_i(u)\in[0,1]$, this forces $F_i(u)=f_i(u)$.

If $\Phi$ is supermodular, then $\PL$ is concave by~\eqref{L3}, and so are the potential term and $\cP$.
Conversely, following~\cite[Proposition~3.6]{Bach13}, a constant guide $F\equiv z\in[0,1]^d$ has linear score zero, because $\E a(U_i)=0$, so $\cP(z)=\PL(z)$.
For sets $S$ and $T$, the layers of $z=\frac12(\ind_S+\ind_T)$ are $S\cup T$ for $s<\frac12$ and $S\cap T$ for $s\ge\frac12$, so by~\eqref{L1} and~\eqref{L2}, concavity of $\cP$ on constant guides gives
\[
\tfrac12\Phi(S\cup T)+\tfrac12\Phi(S\cap T)=\PL(z) \ge\tfrac12\PL(\ind_S)+\tfrac12\PL(\ind_T)=\tfrac12\Phi(S)+\tfrac12\Phi(T),
\]
which is supermodularity.
\end{proof}

\paragraph{Averaging Boolean guides.}
When $\Phi$ is supermodular, averaging Boolean guides cannot increase their average objective gap.
The guides constructed in \cref{sec:guides} use this property.
Let $\xi$ be independent of $U$, and let $X$ be a random Boolean vector with each $X_i$ a function of $(U_{\le i},\xi)$.
The average $g(U)=\E[X\mid U]$ is a guide, and averaging preserves the linear score.
Applying~\eqref{L5} conditionally on $U$, then taking expectations, gives $\E\PL(g(U))\ge\E\Phi(X)$.
Thus
\begin{equation}\label{lat:averaged}
\cP(g)\ge\E_\xi\,\cP\bigl(X(\cdot,\xi)\bigr).
\end{equation}
By~\eqref{lat:descentdeficit} for each fixed $\xi$, the objective gap of $g$ is therefore at most the average over $\xi$ of the rearrangement losses and Bernoulli relative entropies of the Boolean guide $X(\cdot,\xi)$.
In ordinary resampling and in threshold smoothing (\cref{sec:guides}), each $X(\cdot,\xi)$ is the threshold descent of a branching rule, so its rearrangement losses vanish.
When $X\sim\mu$ under the joint law of $(U,\xi)$, as in both constructions, the potential inequality above and $\Hs(\mu)+\E_\mu\Phi=\log Z_\Phi$ give
\[
0\le\log Z_\Phi-\cP(g) \le\Hs(\mu)-\sum_{i=1}^d\E[a(U_i)g_i(U)].
\]

\paragraph{Sampling from a guide.}
Every guide $F$ also defines a sampler, for every potential: $X=\ind\{F(U)>U_0\}$ has a law $\nu$ with $\mathrm{KL}(\nu\Vert\mu)\le\log Z_\Phi-\cP(F)$, by~\eqref{lat:descentdeficit} for each layer, \eqref{lat:layers}, the chain rule and the convexity of relative entropy.

For the coarse objective gaps used here, this bound does not guarantee small total variation distance.
We therefore use \cref{sec:coarseoracle} to turn lower estimates into samples through the tree walk.
\Cref{sec:estimatorconstruction} optimizes a finite-dimensional version of $\cP$ from samples and turns the value it finds into a lower estimate of $\log Z_\Phi$.

\section{Constructing the coarse estimator}\label{sec:estimatorconstruction}

We turn the variational formulation developed in the previous section (\cref{lat:variational}) into an algorithm for the coarse lower estimates~\eqref{conv:oracle} required by our counting and sampling algorithms.
We first use the Hoeffding decomposition to measure interactions among the independent inputs (\cref{sec:hoeffding}).
We then discretize the inputs and restrict to low interaction degree to obtain a finite-dimensional concave optimization problem (\cref{sec:genericestimator}).
The accuracy guarantee assumes the existence of a nearly optimal guide with small high-degree tails.
Such a guide is used only to show that the search space contains a good candidate; the algorithm optimizes and validates candidates using independent uniform inputs and evaluations of the weight $W$.
\Cref{sec:guides} constructs guides satisfying the required bounds.

Throughout this section, $W$ is log-supermodular and satisfies the evaluator and value-length assumptions of \cref{thm:lsm}.
Its normalized potential $\Phi=\log(W/W(0))$ is supermodular, with $Z_\Phi=Z_W/W(0)$ and $\mu=\mu_W$.
The value-length bound gives $2^{-B}<W(x)<2^B$, so every marginal difference of $\Phi$ has absolute value at most $\Lambda=2B$.
Every residual $W_v$ likewise has a supermodular normalized potential $\Phi_v=\log(W_v/W_v(0))$.
Its marginal differences are those of $\log W$ with the prefix fixed, so the same bound $\Lambda$ applies.

\subsection{Interaction degree}\label{sec:hoeffding}

Let $\phi\in L^2((0,1)^d)$ be a function of independent uniform inputs.
For $j\in[d]$ let $E_j$ integrate out coordinate $j$, and for $S\subseteq[d]$ put
\[
\phi_S=\prod_{j\in S}(I-E_j)\prod_{j\notin S}E_j\,\phi ,
\]
where $I$ is the identity.
These are the components of the Hoeffding decomposition $\phi=\sum_S\phi_S$~\cite{Hoeffding48,ES81}, also called the Efron--Stein or functional ANOVA decomposition, and presented as the orthogonal decomposition in~\cite[\S8.3]{ODo14}.
The component $\phi_S$ depends only on the coordinates in $S$ and has mean zero in each of them, and distinct components are orthogonal.

We call $\|\phi_S\|_2^2$ the \emph{interaction energy} on $S$.
For an integer $D\ge0$, the function $\phi$ has \emph{interaction degree at most $D$} if $\phi_S=0$ whenever $|S|>D$.
Write $\Pi_{>D}\phi=\sum_{|S|>D}\phi_S$ for its \emph{degree-$D$ tail}, and $\Pi_{\le D}=I-\Pi_{>D}$.
By orthogonality, $\Pi_{\le D}\phi$ is the best $L^2$ approximation of interaction degree at most $D$, with squared error $\|\Pi_{>D}\phi\|_2^2=\sum_{|S|>D}\|\phi_S\|_2^2$.
These degrees and squared norms correspond to Fourier degree, weight and high-degree weight on the Boolean cube and finite product spaces~\cite[Definitions~1.17, 8.32 and~8.38 and Proposition~8.33]{ODo14}.
Interaction degree counts the input coordinates in a component, not powers of their values or their binary digits.
In \cref{ex:path}, the second row has a nonzero component on $\{1,2\}$.

To represent low-degree functions with finitely many coefficients, we partition each input interval into finitely many bins.
Averaging a function over each product of bins gives a function of the bin indices alone.
This averaging does not increase high-degree tail norms.
Indeed, any linear operator $T$ that commutes with every $E_j$ satisfies $(T\phi)_S=T\phi_S$ for every $S$.
If $T$ is also an $L^2$ contraction, then $\|\Pi_{>D}T\phi\|_2\le\|\Pi_{>D}\phi\|_2$.
Bin averaging has both properties (\cref{sec:witness}).
We will also use this observation for the \emph{noise operator}, which averages a function over independent choices to keep or resample each input (\cref{sec:resampling}).

\subsection{From guides to coarse estimates}\label{sec:genericestimator}

Fix $d\ge2$, $B\ge1$ and an integer $0\le D\le d$, put $\Lambda=2B$, and let $v$ be a node of the self-reduction tree of $W$ whose residual has $k\ge1$ free coordinates, identified with positions $1,\ldots,k$ in the inherited order.
Write $\Phi=\Phi_v$.
The guide and its interaction decomposition use these $k$ inputs.
We allocate the additive slack $1/8$ as $1/32$ for binning and truncation, $1/16$ for optimization, and $1/32$ for validation.
The following parameters depend only on the original dimension $d$ and the bound $B$, uniformly over all residuals:
\begin{equation}\label{alg:parameters}
\begin{gathered}
s=\min\{j\ge0:2^j\ge64d\},\qquad m=4^s,\qquad A_0=\log m+1,\\
G_0=A_0+2\Lambda,\qquad \kappa=\min\{j\ge0:2^j\ge64dG_0\}.
\end{gathered}
\end{equation}
Both minima are over nonnegative integers.
Thus $(64d)^2\le m<(128d)^2$ and $\kappa=O(\log(d+B))$.
Here $m$ is the number of bins per input, $A_0$ bounds the bin scores (\cref{alg:binbounds}), $G_0$ bounds the absolute slope of the objective in each row (\cref{alg:soundness}), and $e^{-\kappa}\le1/(64dG_0)$.
The auxiliary parameter $\kappa$ will be used in \cref{sec:guides} to choose $D$ so that the required tail bound holds.

Suppose that there is a guide $g$ on $k$ inputs for which
\begin{equation}\label{alg:guidehyp}
\log Z_\Phi-\cP(g)\le b_*,\qquad \|\Pi_{>D}g_i\|_2\le\frac1{128dG_0} \quad(1\le i\le k).
\end{equation}
Here $b_*\ge0$ is a known bound.
The guide $g$ may depend on the residual, while $b_*$ and $D$ are common to all residuals.
The first inequality controls the achievable additive error; the second ensures that low-degree truncation loses at most $1/128$ of objective value (\cref{alg:witness}).
We call such a guide a \emph{witness}: it is used only to prove that a good coefficient vector exists, and the algorithm does not evaluate it.

\begin{theorem}[Estimator from a guide]\label{alg:generic}
If~\eqref{alg:guidehyp} holds for every residual, then for every node $v$ and $0<\delta\le1/2$ a randomized algorithm returns $A_v\in[0,d(1+2B)]$ such that
\[
\Pr\bigl[\log Z_{\Phi_v}-b_*-1/8\le A_v\le\log Z_{\Phi_v}\bigr] \ge1-\delta .
\]
At leaves it returns $0$.
Every execution costs at most
\begin{equation}\label{alg:genericcost}
\exp\bigl(O((D+1)\log d)\bigr) \poly(d,B,T_W,\log(1/\delta))
\end{equation}
operations and uniform draws.
\end{theorem}

To prove \cref{alg:generic}, we construct a concave objective bounded above by $\log Z_\Phi$, show that its low-degree search space contains a good candidate, and optimize and validate candidate values using independent samples.
Throughout the proof, $v$, $k$, $\Phi=\Phi_v$ and $\Lambda=2B$ are as above.

\paragraph{A sound concave objective.}\label{sec:bins}
Divide each input interval into $m$ equal bins and put $Y_i=\floor{mU_i}\in\{0,\ldots,m-1\}$ for $1\le i\le k$.
The variables $Y_i$ are independent and uniform.
Low-degree truncation need not preserve the range $[0,1]$.
For optimization, a \emph{candidate} is therefore a vector $u=(u_1,\ldots,u_k)$ whose $i$th row is a real-valued function $u_i(Y_1,\ldots,Y_i)$.
Clipping its rows to $[0,1]$ produces a guide.

For a function of the bins, replacing $a(U_i)$ by its conditional mean preserves the linear score.
The score of bin $j$ is
\[
\bar a_j=m\int_{j/m}^{(j+1)/m}a(t)\,dt =m\bigl[\Hs(j/m)-\Hs((j+1)/m)\bigr],
\]
since $\Hs'=-a$ on $(0,1)$.
The following bounds control the size of these scores and the score error when a guide is averaged over bins.

\begin{lemma}[Bin scores]\label{alg:binbounds}
The bin scores satisfy $\max_j|\bar a_j|\le A_0$ and $\E|a(U_1)-\bar a_{Y_1}|\le(2\log m+6)/m$.
\end{lemma}
\begin{proof}
Since $a(1-t)=-a(t)$, we have $\bar a_{m-1-j}=-\bar a_j$.
The interior bins lie in $[1/m,1-1/m]$, where $|a|\le\log(m-1)$.
For the first bin, $m\int_0^{1/m}\log t\,dt=-\log m-1$ and $0\le-m\int_0^{1/m}\log(1-t)\,dt\le-\log(1-1/m)\le2/m$, so $\bar a_0\in[-A_0,-A_0+2/m]$.
This proves the first bound.

An increasing function $\phi$ with mean $\bar\phi$ on an interval of length $1/m$ satisfies $\int|\phi-\bar\phi|=2\int(\phi-\bar\phi)_+\le\frac2m(\sup\phi-\bar\phi)$, and likewise with $\bar\phi-\inf\phi$; hence $\int|\phi-\bar\phi|$ is at most $1/m$ times the oscillation of $\phi$.
The oscillations of $a$ over the interior bins sum to $a(1-1/m)-a(1/m)=2\log(m-1)$.
For the boundary bins, we use the one-sided bound.
On the first bin $\sup a-\bar a_0\le-\log(m-1)+\log m+1\le1+2/m$, and the last bin is symmetric.
Summing, $\E|a(U_1)-\bar a_{Y_1}|\le(2\log(m-1)+4+8/m)/m\le(2\log m+6)/m$.
\end{proof}

Substituting the clipped candidate into $\cP$ gives a lower bound, but need not preserve concavity.
We therefore extend the bin scores concavely and subtract a penalty that controls the change in the potential under clipping.
For $t\in\R$ let $\psi_t(y)=t\min\{y,1\}$ if $t\ge0$ and $\psi_t(y)=t\max\{y,0\}$ if $t<0$.
This function is concave and $|t|$-Lipschitz, equals $ty$ on $[0,1]$, and satisfies $\psi_t(y)\le t\clip(y)$, where $\clip(y)=\min\{1,\max\{0,y\}\}$; we write $\dist(y,[0,1])=|y-\clip(y)|$ for the distance to the interval.
For a candidate $u$ define the \emph{penalized objective}
\begin{equation}\label{alg:objective}
J_v(u)=\sum_{i=1}^k\E\psi_{\bar a_{Y_i}}\bigl(u_i(Y)\bigr) +\E\PL(u(Y)) -\Lambda\sum_{i=1}^k\E\dist\bigl(u_i(Y),[0,1]\bigr).
\end{equation}
The objective is an expectation under the fixed product law of the bins; its integrand uses only candidate values and evaluations of $W$.

\begin{lemma}[Soundness for every candidate]\label{alg:soundness}
The objective $J_v$ is concave.
Every candidate $u$ satisfies $J_v(u)\le\log Z_\Phi$, and any two candidates satisfy $|J_v(u)-J_v(u')|\le G_0\sum_i\E|u_i-u'_i|$.
\end{lemma}
\begin{proof}
Concavity holds termwise: $\psi_t$, $\PL$ and minus the distance to an interval are concave, and $u\mapsto u_i(Y)$ is linear.
Let $F=\clip(u)$, coordinatewise; it is a guide.
Then $\psi_t(u_i)\le tF_i$ for every $t$, and~\eqref{L4} gives $\PL(u)\le\PL(F)+\Lambda\sum_i\dist(u_i,[0,1])$: the penalty pays for the change of the potential term under clipping.
Hence
\[
J_v(u)\le\sum_i\E[\bar a_{Y_i}F_i]+\E\PL(F)=\cP(F)\le\log Z_\Phi .
\]
The equality holds because $F_i$ is a function of the bins and $\bar a_{Y_i}=\E[a(U_i)\mid Y]$; the last inequality is \cref{lat:variational}.

For the Lipschitz bound, the term $\psi_{\bar a_{Y_i}}(u_i)$ of row $i$ is $A_0$-Lipschitz by \cref{alg:binbounds}, and the potential term and the penalty are $\Lambda$-Lipschitz in each row.
\end{proof}

\paragraph{A good low-degree candidate exists.}\label{sec:witness}
We now represent low-degree candidates by finitely many bounded coefficients and show that one preserves nearly the witness's objective value.
We use Walsh characters because their orthogonality controls truncation error, while their values in $\{-1,1\}$ ensure that $[0,1]$-valued functions of the bins have coefficients in $[-1,1]$.
Since $m=2^{2s}$, identify each bin index $y\in\{0,\ldots,m-1\}$ with its vector of $2s$ binary digits.
For $\beta\in\{0,1\}^{2s}$ put $\chi_\beta(y)=(-1)^{\beta\cdot y}$, where $\beta\cdot y$ is the inner product of the digit vectors, and for $\beta=(\beta_1,\ldots,\beta_k)$ put $\chi_\beta(Y)=\prod_j\chi_{\beta_j}(Y_j)$.
These $m^k$ Walsh characters~\cite{ODo14} are an orthonormal basis for the functions of $Y$.
The \emph{support} of $\beta$ is $\{j:\beta_j\ne0\}$.
For a function $\phi$ of the bins write $\hat\phi(\beta)=\E[\phi\,\chi_\beta]$.
Viewed as a function of $U$, the Hoeffding component of $\phi$ on $S$ is $\sum\hat\phi(\beta)\chi_\beta$ over the $\beta$ with support $S$~\cite[Proposition~8.36]{ODo14}.

\begin{lemma}[Low-degree candidate]\label{alg:witness}
There is a candidate $u^*$ such that each row $u_i^*$ is a linear combination, with coefficients in $[-1,1]$, of the Walsh characters whose supports lie in $\{1,\ldots,i\}$ and have at most $D$ elements, and
\[
J_v(u^*)\ge\log Z_\Phi-b_*-\frac1{32}.
\]
\end{lemma}
\begin{proof}
Let $g$ be the guide assumed in~\eqref{alg:guidehyp}, and put $\bar g_i=\E[g_i\mid Y]$.
Conditioning on $Y$ averages the guide over each product of input bins.
Thus $\bar g$ takes values in $[0,1]^k$, and $\bar g_i$ depends only on $Y_1,\ldots,Y_i$.
Its penalty vanishes, and since $\PL$ is concave, $\PL(\bar g(Y))\ge\E[\PL(g(U))\mid Y]$.
Since $g_i\in[0,1]$, \cref{alg:binbounds} and the choice of $m$ give
\[
J_v(\bar g)\ge\sum_i\E[\bar a_{Y_i}g_i]+\E\PL(g) \ge\cP(g)-\frac{k(2\log m+6)}m \ge\log Z_\Phi-b_*-\frac1{64}.
\]
For the last step use~\eqref{alg:guidehyp} and $d(2\log m+6)/m<1/64$: with $d\le2^s/64$ and $s\ge6$ the left-hand side is at most $(4s\log2+6)/(64\cdot2^s)<1/64$.

Now truncate to interaction degree $D$, setting $u_i^*=\Pi_{\le D}\bar g_i$.
Conditioning on the bins acts on each coordinate separately, commutes with every $E_j$ and does not increase norms, so by \cref{sec:hoeffding} $\Pi_{>D}\bar g_i=\E[\Pi_{>D}g_i\mid Y]$, and~\eqref{alg:guidehyp} gives $\|\Pi_{>D}\bar g_i\|_2\le1/(128dG_0)$.
By Cauchy--Schwarz, $\E|\bar g_i-u_i^*|\le1/(128dG_0)$.
Summing over the $k\le d$ rows, \cref{alg:soundness} turns this into an objective loss of at most $1/128$.
Together with the preceding bound, this gives the claimed objective value.
In the Walsh basis $u_i^*$ keeps exactly the characters of $\bar g_i$ whose supports have at most $D$ elements.
Their coefficients $\E[\bar g_i\chi_\beta]$ lie in $[-1,1]$ because $0\le\bar g_i\le1$, and they vanish unless the support lies in $\{1,\ldots,i\}$, since $\bar g_i$ does not depend on later bins.
\end{proof}

\paragraph{Stochastic optimization.}\label{sec:cube}
For row $i$ let $\mathcal F_i$ be the set of $\beta$ whose support lies in $\{1,\ldots,i\}$ and has at most $D$ elements.
Put
\begin{equation}\label{alg:featuredimension}
M_i=|\mathcal F_i|=\sum_{j=0}^{\min(D,i)}\binom ij(m-1)^j,\qquad M=\max_iM_i,\qquad P=\sum_{i=1}^kM_i\le kM.
\end{equation}
A coefficient vector $\theta\in\mathcal C=[-1,1]^P$ (unrelated to the branching rules of \cref{sec:treeprinciple}) defines the candidate $u^\theta_i=\sum_{\beta\in\mathcal F_i}\theta_{i,\beta}\chi_\beta$, and we write $J(\theta)=J_v(u^\theta)$.
By \cref{alg:witness}, some $\theta^\circ\in\mathcal C$ has $J(\theta^\circ)\ge\log Z_\Phi-b_*-1/32$.
Since $M\le(1+dm)^D$ and $m=\poly(d)$,
\begin{equation}\label{alg:dimcost}
\log P\le\log d+D\log(1+dm)=O\bigl((D+1)\log d\bigr).
\end{equation}

A fresh sample of the bins allows us to evaluate a sample objective and a stochastic supergradient.
Generating the sample uses $k$ uniform draws and $O(k\log m)$ comparisons.
At a sample every character is a sign, so $|u^\theta_i(Y)|\le M$.
Dropping the expectations in~\eqref{alg:objective} gives the sample objective $\widehat J_Y(\theta)$, which is concave and piecewise linear in $\theta$.
For every fixed $\theta$, $\E\widehat J_Y(\theta)=J(\theta)$.
The potential term of $\widehat J_Y(\theta)$ is evaluated by sorting the $k$ values $u^\theta_i(Y)$ and applying~\eqref{lat:lovasz} to $k+1$ values of $W$.
Choose a supergradient $\zeta_Y(\theta)$ from the chain increments of that sorting~\eqref{L3}, with slope zero at every kink of $\psi_t$ and of the penalty.
The slope of each row then has magnitude at most $G_0$.
The $(i,\beta)$ coordinate of $\zeta_Y(\theta)$ is the chosen row-$i$ slope multiplied by $\chi_\beta(Y)$.
Put $G_1=kG_0$.
Since $\PL(0)=0$, the sample objective satisfies $|\widehat J_Y(\theta)|\le G_0\sum_i|u^\theta_i(Y)|\le MG_1$, and
\begin{equation}\label{alg:samplebounds}
\|\zeta_Y(\theta)\|_2^2\le\sum_iM_iG_0^2\le MG_1^2 .
\end{equation}
Averaging the supergradient inequality gives a supergradient $\E\zeta_Y(\theta)$ of $J$ at fixed $\theta$.

We use projected stochastic supergradient ascent~\cite[\S2.2]{NJLS09} with target expected optimization gap $\gamma=1/64$.
Set $\alpha=\gamma/(MG_1^2)$ and $n_{\rm it}=\ceil{PMG_1^2/\gamma^2}$.
Initialize $\theta_1=0$ and iterate
\begin{equation}\label{alg:update}
\theta_{j+1}=\Pi_{\mathcal C}\bigl(\theta_j+\alpha\,\zeta_{Y^{(j)}}(\theta_j)\bigr), \qquad j=1,\ldots,n_{\rm it},
\end{equation}
with a fresh sample $Y^{(j)}$ each time; the projection $\Pi_{\mathcal C}$ clips each coordinate to $[-1,1]$.
Return
\[
\bar\theta=n_{\rm it}^{-1}\sum_{j=1}^{n_{\rm it}}\theta_j.
\]

If $\theta^*$ maximizes $J$ on $\mathcal C$, then projection does not increase the distance to $\theta^*$, and the conditional supergradient inequality gives
\[
\E\|\theta_{j+1}-\theta^*\|_2^2\le\E\|\theta_j-\theta^*\|_2^2 -2\alpha\,\E[J(\theta^*)-J(\theta_j)]+\alpha^2MG_1^2 .
\]
Summing over $j$, using $\|\theta_1-\theta^*\|_2^2\le P$ and concavity of $J$ at $\bar\theta$,
\begin{equation}\label{alg:optimizerbound}
\E[J(\theta^*)-J(\bar\theta)]\le\frac P{2\alpha n_{\rm it}}+\frac{\alpha MG_1^2}2\le\gamma .
\end{equation}
By Markov's inequality, $J(\bar\theta)\ge J(\theta^*)-4\gamma$ with probability at least $3/4$.
The finite-precision implementation and its rounding analysis are given in \cref{sec:finitealgorithm}.

\paragraph{Validation.}
For a requested failure probability $0<\delta\le1/2$, run the ascent independently $n_{\rm att}=\ceil{\log_2(4/\delta)}$ times to amplify the probability of finding a good candidate.
We estimate each returned candidate's objective value using fresh samples, since the candidate depends on the samples used during ascent.
Subtracting the validation error margin then gives a lower estimate of $\log Z_\Phi$ whenever validation is accurate.
For each candidate, average $\widehat J_Y$ over
\begin{equation}\label{alg:validation}
n_{\rm val}=\ceil{2M^2G_1^2\gamma^{-2}\log(8n_{\rm att}/\delta)}
\end{equation}
fresh samples.
Conditional on the candidate, Hoeffding's inequality~\cite{Hoe63} shows that this average differs from its objective value by more than $\gamma$ with probability at most $\delta/(4n_{\rm att})$.

\begin{proof}[Proof of \cref{alg:generic}]
At a leaf return $0$.
Otherwise compute the parameters~\eqref{alg:parameters} by comparisons, and the $m$ bin scores from the $m+1$ values $\Hs(j/m)$.
Run $n_{\rm att}$ ascent and validation attempts as above, writing $\bar\theta^{(\rho)}$ and $\widehat J^{(\rho)}$ for the averaged iterate and validation average in attempt $\rho$.
Return
\[
A_v=\min\Bigl\{k(1+\Lambda),\max\bigl\{0,\max_\rho(\widehat J^{(\rho)}-\gamma)\bigr\}\Bigr\}.
\]

By a union bound, with probability at least $1-\delta/4$ every validation is accurate within $\gamma$, and with probability at least $1-4^{-n_{\rm att}}\ge1-\delta/4$ some attempt has $J(\bar\theta^{(\rho)})\ge J(\theta^*)-4\gamma$.
On both events every $\widehat J^{(\rho)}-\gamma$ is at most $J(\bar\theta^{(\rho)})\le\log Z_\Phi$ by \cref{alg:soundness}, and the best one is at least
\[
J(\theta^*)-6\gamma\ge J(\theta^\circ)-6\gamma \ge\log Z_\Phi-b_*-\frac1{32}-\frac6{64} =\log Z_\Phi-b_*-\frac18 .
\]
The clamps keep $A_v$ in $[0,k(1+\Lambda)]\subseteq[0,d(1+2B)]$, since $k\le d$.
They are harmless because $0\le\log Z_\Phi\le k(\log2+\Lambda)<k(1+\Lambda)$: the configuration $0$ contributes $e^{\Phi(0)}=1$, and $\Phi(x)\le k\Lambda$ by telescoping from $0$.

For the cost, the characters $\chi_\beta$ with $\beta\in\mathcal F_i$ are listed by choosing a support of at most $D$ positions and nonzero digit patterns on it, and a character is evaluated in $O(1+Ds)$ operations.
One ascent step or one validation sample costs $\poly(d)P$ operations and $k+1$ evaluations of $W$.
There are $n_{\rm att}n_{\rm it}$ steps and $n_{\rm att}n_{\rm val}$ validation samples, all fixed in advance, and $G_1=\poly(d,B)$.
By~\eqref{alg:dimcost} and $M,P\le\exp(O((D+1)\log d))$, the total is~\eqref{alg:genericcost} on every execution.
\end{proof}

\section{Low-degree approximation and threshold smoothing}\label{sec:guides}

We construct nearly optimal guides with small high-degree tails, providing the witnesses required by the estimator of \cref{alg:generic}.
These guides are obtained by averaging random perturbations of the exact guide and serve only to certify that the search space contains a good candidate.
We first analyze ordinary resampling, which gives the square-root baseline (\cref{sec:resampling}).
We then introduce threshold smoothing, which improves the objective-loss bound from linear to quadratic in the perturbation parameter while preserving comparable high-degree decay (\cref{sec:smoothing}).
This improved tradeoff completes the proof of \cref{alg:coarsetheorem} and yields the cube-root running-time exponent.

Throughout this section, $\Phi$ is a supermodular potential on $\{0,1\}^d$, normalized by $\Phi(0)=0$, and $\mu=e^\Phi/Z_\Phi$.
We use the exact guide $f$ of \eqref{lat:encoder}, the variational objective $\cP$, and the Hoeffding decomposition of \cref{sec:hoeffding}.
The constructions work in any fixed coordinate order and apply, in particular, to every pinned residual from \cref{sec:coarseoracle}.

\subsection{Ordinary resampling and the square-root baseline}\label{sec:resampling}\label{sec:baseline}

The simplest way to shrink the high-degree components of the exact guide~\eqref{lat:encoder} is to resample its inputs: for $0<\tau\le1/2$, let $g=T_\tau f$, where $T_\tau=\prod_{j=1}^d((1-\tau)I+\tau E_j)$ is the noise operator.
O'Donnell~\cite[Definitions~8.26--8.27, Proposition~8.28 and \S8.3]{ODo14} treats it on finite product spaces, with the correlation $1-\tau$ as its parameter.

\begin{lemma}[Ordinary resampling]\label{guide:resampling}
For every $0<\tau\le1/2$, $g=T_\tau f$ is a guide, and
\[
0\le\log Z_\Phi-\cP(g)\le\tau\Hs(\mu)\le d\tau\log2, \qquad \|\Pi_{>D}g_i\|_2\le\tfrac12(1-\tau)^{D+1}
\]
for every row $i$ and integer $D\ge0$.
\end{lemma}
\begin{proof}
The operator is positive, fixes constants, and preserves the uniform product measure.
Because $f_i$ depends only on the first $i$ inputs, so does $g_i$.
Sample a noisy copy $V$ of $U$ by independently keeping each coordinate with probability $1-\tau$ and otherwise drawing a fresh uniform value.
Then $g(U)=\E[f(V)\mid U]$, while $V$ is uniform.
As in~\eqref{lat:averaged}, \eqref{L5} applied given $U$ gives $\E\PL(g)\ge\E\Phi(f(V))=\E_\mu\Phi$.
Consider the linear score of row $i$.
If $U_i$ is resampled, the row contributes zero, since $\E a(U_i)=0$.
If $U_i$ is kept, the noisy prefix $V_{<i}$ has the uniform product law and is independent of $U_i$; hence
\[
\E[a(U_i)g_i(U)]=(1-\tau)\E[a(U_i)f_i(U)] .
\]
By \cref{lat:exactencoder}, the sum of the latter expectations is $\Hs(\mu)$.
This proves the objective-gap bound using \cref{lat:variational}.
Finally $T_\tau$ multiplies the Hoeffding component on $S$ by $(1-\tau)^{|S|}$.
Orthogonality and $\Var(f_i)\le1/4$ give $\|\Pi_{>D}g_i\|_2^2\le\frac14(1-\tau)^{2(D+1)}$.
\end{proof}

\begin{corollary}[Square-root baseline]\label{alg:baseline}
Ordinary resampling and the estimator of \cref{alg:generic} give the counting and sampling guarantees of \cref{thm:lsm} within the bound
\[
\exp\Bigl(O\bigl(\min\{d,\sqrt{d\log d\,\log(d+B)}\}\bigr)\Bigr)
\]
times the same polynomial factors.
For bipartite independent sets this gives $\exp(O(\sqrt n\log n))$ times the polynomial factors in \cref{thm:main}.
\end{corollary}

\begin{proof}
For $0<\tau\le1/2$, take $D=\min\{d,\ceil{\kappa/\tau}\}$, with the parameters of~\eqref{alg:parameters}.
Apply \cref{guide:resampling} to any residual with $k\le d$ free coordinates.
Its objective gap is at most $d\tau\log2$.
If $D<d$, the degree-$D$ tail of each row has norm at most $\frac12(1-\tau)^{D+1}\le\frac12e^{-\kappa}\le1/(128dG_0)$; if $D=d$ the tail vanishes.
Thus \cref{alg:generic} supplies, at every node, a one-sided estimate with additive error bound $b=d\tau\log2+1/8$ and cost
\[
\exp\bigl(O(\tau^{-1}\log d\,\log(d+B))\bigr) \poly(d,B,T_W,\log(1/\delta)).
\]
The conversion of \cref{conv:conversion} therefore has exponential cost $\exp(O(d\tau+\ell_d\ell_B/\tau))$, with $\ell_d$ and $\ell_B$ as in the proof of \cref{thm:lsm}.

If $\ell_d\ell_B<d/4$, choose the largest $\tau=2^{-j}$ with integer $j\ge1$ for which $d\tau^2\le\ell_d\ell_B$.
Then both $d\tau$ and $\ell_d\ell_B/\tau$ are $O(\sqrt{d\ell_d\ell_B})$.
If $\ell_d\ell_B\ge d/4$, exact enumeration instead costs $2^d\poly(d,T_W)$, and $d=O(\sqrt{d\ell_d\ell_B})$.
Taking the cheaper of the resulting algorithm and enumeration gives the stated minimum.
Finally, $\ell_d\ell_B=O(\log d\,\log(d+B))$; for \BIS\ use $d\le n$, $B\le n+1$, and $n\ge2$.
\end{proof}

\subsection{Threshold smoothing and the two guarantees}\label{sec:smoothing}

Resampling loses linearly in $\tau$, because a resampled input contributes nothing to the linear score.
Threshold smoothing keeps every input and shifts its threshold instead, which loses only quadratically while reducing the variance of the averaged indicator at each threshold by a fraction linear in $w$.

\begin{theorem}[Near-optimal guides with small high-degree tails]\label{guide:theorem}
Let $\Phi$ be a normalized supermodular potential on $d$ coordinates.
For every $0<w\le1/2$ there is a guide $g$ such that
\begin{enumerate}[label=(\roman*),ref=\roman*]
\item\label{guide:pressure}
\emph{Near-optimal objective value:} $0\le\log Z_\Phi-\cP(g)\le dw^2/4$, and
\item\label{guide:degree}
\emph{Low-degree approximation:} $\|\Pi_{>D}g_i\|_2\le\frac12(1-w/2)^{(D+1)/2}$ for every row $i$ and every integer $D\ge0$.
\end{enumerate}
\end{theorem}

We construct $g$, in any coordinate order, by smoothing the thresholds of the exact guide.
Item~(\ref{guide:pressure}) controls the additive error in the estimate of $\log Z_\Phi$; item~(\ref{guide:degree}) controls the degree needed for an accurate truncation of $g$, and hence the size of the search space.

\paragraph{The construction.}
Fix a smoothing parameter $0<w\le1/2$.

\begin{definition}[Ordered threshold tree]\label{lat:tree}
A function $\mathcal T:(0,1)^d\to\R$ is an \emph{ordered threshold tree} if it can be computed as follows.
For $j=1,\ldots,d$ in turn, compare $u_j$ with a threshold $c_j\in[0,1]$ that depends only on the outcomes $x_1,\ldots,x_{j-1}$ of the previous comparisons, and record $x_j=\ind\{u_j\ge c_j\}$.
Finally output a real number that depends only on $x_1,\ldots,x_d$.
Equivalently, $\mathcal T$ is computed by a decision tree that queries $u_1,\ldots,u_d$ in this order, each query a threshold comparison.
\end{definition}

Each row $f_i$ of the exact guide is an ordered threshold tree: for $j\le i$ the threshold is $c_j=1-p_j(x_{<j})$, the later comparisons are ignored, and the output is $x_i$.
Earlier inputs influence a later threshold only through the earlier binary decisions.

For $c\in[0,1]$ put $h(c)=wc(1-c)$.
To \emph{smooth} an ordered threshold tree, draw independent fair signs $\sigma_1,\ldots,\sigma_d\in\{-1,1\}$, independent of the uniform inputs, and in the $j$th comparison replace $c_j$ by $c_j+\sigma_j h(c_j)$.
Here $c_j$ is the threshold at the node reached by the \emph{perturbed} comparisons.
Write $\mathcal T^\sigma(U)$ for the resulting output, and define its smoothing by averaging the signs at fixed inputs:
\begin{equation}\label{sm:blur}
g_{\mathcal T}(u)=\E_\sigma \mathcal T^\sigma(u).
\end{equation}
We call this perturbation and averaging operation \emph{threshold smoothing}, and $g_{\mathcal T}$ the \emph{smoothing} of $\mathcal T$; it need not be continuous.
Both shifted thresholds lie in $[0,1]$.
Conditional on the preceding inputs and signs, the probability of taking branch $1$ is
\[
\tfrac12\bigl(1-c_j-h(c_j)\bigr) +\tfrac12\bigl(1-c_j+h(c_j)\bigr)=1-c_j.
\]
By induction on $j$, the perturbed outcomes $(x_1,\ldots,x_d)$, with inputs and signs random, have the same law as the unperturbed ones.
In particular, under the joint law of $(U,\sigma)$, $\mathcal T^\sigma(U)$ has the law of $\mathcal T(U)$, and $\E g_{\mathcal T}=\E \mathcal T$.

For the exact guide, use one perturbed sequential run for all rows: with independent uniform inputs $U_i$ and independent fair signs $\sigma_i$, set
\begin{equation}\label{sm:perturbedencoder}
c_i^\sigma=1-p_i(X^\sigma_{<i}),\qquad X_i^\sigma=\ind\{U_i\ge c_i^\sigma+\sigma_i h(c_i^\sigma)\},\qquad g(U)=\E_\sigma[X^\sigma\mid U].
\end{equation}
Thus $g_i=g_{f_i}$.
Averaging each row with its own independent sign sequence would give the same $g$.
By the calculation above, $X^\sigma\sim\mu$ when inputs and signs are both random; for fixed signs the law is in general different.
The shared run gives one random Boolean vector $X^\sigma$ with conditional mean $g(U)$ and law $\mu$, which the potential bound of \cref{sm:encoder} needs.

\paragraph{The two guarantees.}
\Cref{sm:encoder} bounds the objective gap of the smoothed guide, and \cref{sm:energy} bounds the high-degree tails of the smoothing of any ordered threshold tree.
Together they prove \cref{guide:theorem}.

\begin{proposition}[Near-optimal objective value for the smoothed guide]\label{sm:encoder}
The guide $g$ of~\eqref{sm:perturbedencoder} has the following properties.
\begin{enumerate}[label=(\arabic*),ref=\arabic*]
\item\label{sm:enc1}
$g$ is a guide.
\item\label{sm:enc2}
With $X\sim\mu$, $c_i=1-p_i(X_{<i})$ and $\Delta_w(c)=\Hs(c)-\frac12\Hs(c-h(c))-\frac12\Hs(c+h(c))$,
\begin{equation}\label{sm:lossidentity}
\Hs(\mu)-\sum_{i=1}^d\E[a(U_i)g_i(U)]=\sum_{i=1}^d\E\Delta_w(c_i).
\end{equation}
\item\label{sm:enc3}
For every $c\in[0,1]$, $0\le\Delta_w(c)\le w^2c(1-c)$.
\item\label{sm:enc4}
$\E\PL(g(U))\ge\E_\mu\Phi$.
Consequently
\begin{equation}\label{sm:loss}
0\le\log Z_\Phi-\cP(g)\le w^2\sum_{i=1}^d\E[c_i(1-c_i)]\le\frac{dw^2}4.
\end{equation}
\end{enumerate}
\end{proposition}

\begin{theorem}[Energy identity]\label{sm:energy}
Let $\mathcal T$ be an ordered threshold tree on $d$ coordinates and $g_{\mathcal T}$ its smoothing.
Then
\begin{equation}\label{sm:identity}
\sum_{S\subseteq[d]}\Bigl(1-\frac w2\Bigr)^{-|S|}\|(g_{\mathcal T})_S\|_2^2=\|\mathcal T\|_2^2.
\end{equation}
Consequently $\sum_{S\ne\varnothing}(1-w/2)^{-|S|}\|(g_{\mathcal T})_S\|_2^2=\Var(\mathcal T)$, and for every integer $D\ge0$,
\begin{equation}\label{sm:tail}
\|\Pi_{>D}g_{\mathcal T}\|_2^2\le\Bigl(1-\frac w2\Bigr)^{D+1}\Var(\mathcal T).
\end{equation}
\end{theorem}

\begin{proof}[Proof of \cref{guide:theorem}]
Take $g$ as in~\eqref{sm:perturbedencoder}.
By \cref{sm:encoder} it is a guide satisfying the near-optimal objective value guarantee~(\ref{guide:pressure}).
For the low-degree approximation guarantee~(\ref{guide:degree}), apply~\eqref{sm:tail} to the ordered threshold tree $f_i$, whose variance is at most $1/4$.
\end{proof}

\Cref{guide:theorem} supplies the witness required by \cref{alg:generic}, which completes the main line.

\begin{proof}[Proof of \cref{alg:coarsetheorem}]
Apply the generic estimator of \cref{alg:generic} with $b_*=dw^2/4$ and $D=\min\{d,\ceil{4\kappa/w}\}$.
Apply \cref{guide:theorem} to the normalized potential $\Phi_v$ of each residual, which is supermodular with $k\le d$ free coordinates; its objective gap is at most $kw^2/4\le b_*$.
If $D<d$, item~(\ref{guide:degree}) of \cref{guide:theorem} and $1-w/2\le e^{-w/2}$ give $\|\Pi_{>D}g_i\|_2\le\frac12e^{-\kappa}\le1/(128dG_0)$; if $D=d$ the tail vanishes.
The generic cost is~\eqref{alg:genericcost}, and $D=O(w^{-1}\log(d+B))$.
\end{proof}

\paragraph{Quadratic objective gap.}
It remains to prove \cref{sm:encoder,sm:energy}.
The objective gap uses the one-bit identity~\eqref{lat:bernoulli} for the linear score and the averaging argument of~\eqref{lat:averaged} for the potential term.

\begin{proof}[Proof of \cref{sm:encoder}]
(\ref{sm:enc1}) Each $X_i^\sigma$ is Boolean and depends only on $U_{\le i},\sigma_{\le i}$, so its average lies in $[0,1]$ and depends only on $U_{\le i}$.

(\ref{sm:enc2}) Conditional on $U_{<i},\sigma_{<i}$, $c_i^\sigma$ is fixed and $(U_i,\sigma_i)$ is fresh.
By~\eqref{lat:bernoulli},
\[
\E[a(U_i)X_i^\sigma\mid U_{<i},\sigma_{<i}] =\tfrac12\Hs(c_i^\sigma-h(c_i^\sigma)) +\tfrac12\Hs(c_i^\sigma+h(c_i^\sigma)).
\]
Since $X^\sigma\sim\mu$, $c_i^\sigma$ has the law of $c_i$.
Also $\E[a(U_i)g_i(U)]=\E[a(U_i)X_i^\sigma]$, and $\Hs(\mu)=\sum_i\E\Hs(c_i)$ by \cref{lat:exactencoder}.
Subtracting proves~\eqref{sm:lossidentity}.

(\ref{sm:enc3}) At $c=0,1$ the loss is zero.
For $0<c<1$, put $h=h(c)$.
The two shifted thresholds $c\pm h$ have mean $c$, so expanding Bernoulli relative entropy gives
\begin{equation}\label{sm:kl}
\Delta_w(c)=\tfrac12\mathrm{KL}\bigl(\mathrm{Ber}(c-h)\Vert\mathrm{Ber}(c)\bigr) +\tfrac12\mathrm{KL}\bigl(\mathrm{Ber}(c+h)\Vert\mathrm{Ber}(c)\bigr).
\end{equation}
The inequality $\log t\le t-1$ yields
\[
\mathrm{KL}\bigl(\mathrm{Ber}(q)\Vert\mathrm{Ber}(c)\bigr) \le\frac{(q-c)^2}{c(1-c)}\qquad(0\le q\le1),
\]
by applying it to the two probability ratios.
Hence $0\le\Delta_w(c)\le h^2/[c(1-c)]=w^2c(1-c)$.

(\ref{sm:enc4}) Given $U$, the Boolean vector $X^\sigma$ has mean $g(U)$, so~\eqref{L5} gives $\E[\Phi(X^\sigma)\mid U]\le\PL(g(U))$, as in~\eqref{lat:averaged}.
Averaging and using $X^\sigma\sim\mu$ proves the potential inequality.
Together with items~(\ref{sm:enc2}) and~(\ref{sm:enc3}) this gives $\cP(g)\ge\Hs(\mu)+\E_\mu\Phi-w^2\sum_i\E[c_i(1-c_i)]$.
Now $\Hs(\mu)+\E_\mu\Phi=\log Z_\Phi$, the lower bound in~\eqref{sm:loss} is \cref{lat:variational}, and $c(1-c)\le1/4$.
\end{proof}

\paragraph{Geometric decay of high-degree tails.}
We prove \cref{sm:energy} by induction on $d$, using only the mean and variance of a single smoothed indicator.

Let $\iota_c=\ind_{[c,1)}$ be the threshold indicator of \cref{sec:descents}.
The smoothing of a single comparison is
\begin{equation}\label{sm:cutblur}
\phi_c(u)=\tfrac12 \iota_{c-h(c)}(u)+\tfrac12 \iota_{c+h(c)}(u).
\end{equation}

\begin{lemma}[Moments of a smoothed indicator]\label{sm:kernel}
For every $c\in[0,1]$, with $p=1-c$,
\begin{equation}\label{sm:cutvariance}
\E\phi_c=p,\qquad \Var(\phi_c)=\Bigl(1-\frac w2\Bigr)p(1-p).
\end{equation}
\end{lemma}
\begin{proof}
Write $h=h(c)$.
The two indicators in~\eqref{sm:cutblur} are nested, with means $p+h$ and $p-h$, so $\E\phi_c=p$ and $\E\phi_c^2=\frac14(p+h)+\frac34(p-h)=p-h/2$.
Subtracting $p^2$ gives $p(1-p)-h/2=(1-w/2)c(1-c)$.
This also holds at $c=0,1$, where $h=0$.
\end{proof}

\begin{proof}[Proof of \cref{sm:energy}]
Put $\rho=1-w/2$.
On any product of input coordinates, use the squared norm $\|H\|_\rho^2=\sum_S\rho^{-|S|}\|H_S\|_2^2$, with Hoeffding components as in \cref{sec:hoeffding}.
We prove $\|g_{\mathcal T}\|_\rho^2=\|\mathcal T\|_2^2$ by induction on $d$; for $d=0$ the tree and its smoothing are the same constant.

For $d\ge1$, let $c$ be the first threshold, put $p=1-c$, and let $\mathcal T^{(1)}$ and $\mathcal T^{(0)}$ be the two remaining ordered threshold trees, with smoothings $g^{(1)}$ and $g^{(0)}$.
Writing $u'=(u_2,\ldots,u_d)$, independence of the signs gives
\begin{equation}\label{sm:recursion}
g_{\mathcal T}(u)=\phi_c(u_1)g^{(1)}(u')+(1-\phi_c(u_1))g^{(0)}(u') =A(u')+\bigl(\phi_c(u_1)-p\bigr)H(u'),
\end{equation}
where $A=pg^{(1)}+(1-p)g^{(0)}$ and $H=g^{(1)}-g^{(0)}$.
The first summand has no interaction involving coordinate $1$.
In the second, the factor $\phi_c-p$ is centered, so a component of $H$ on $S\subseteq\{2,\ldots,d\}$ produces a component on $S\cup\{1\}$.
Orthogonality and \cref{sm:kernel} therefore give
\begin{align*}
\|g_{\mathcal T}\|_\rho^2 &=\|A\|_\rho^2+\rho^{-1}\Var(\phi_c)\|H\|_\rho^2\\
&=\|pg^{(1)}+(1-p)g^{(0)}\|_\rho^2+p(1-p)\|g^{(1)}-g^{(0)}\|_\rho^2\\
&=p\|g^{(1)}\|_\rho^2+(1-p)\|g^{(0)}\|_\rho^2.
\end{align*}
The last equality is the usual quadratic identity in an inner-product norm.
By the induction hypothesis, the last line equals $p\|\mathcal T^{(1)}\|_2^2+(1-p)\|\mathcal T^{(0)}\|_2^2$, which is $\|\mathcal T\|_2^2$ because $\mathcal T$ follows $\mathcal T^{(1)}$ when $u_1\ge c$ and $\mathcal T^{(0)}$ otherwise.
This proves~\eqref{sm:identity}.

The constant component is $\E g_{\mathcal T}=\E \mathcal T$.
Subtracting its square gives the variance identity; for the tail, use $(1-w/2)^{|S|}\le(1-w/2)^{D+1}$ when $|S|>D$.
\end{proof}

The factor $\rho^{-1}$ for the added coordinate $1$ exactly offsets the variance that the smoothing removes from the indicator, so the weighted energy passes down the tree without loss.
This is an aggregate identity: an individual component's squared norm need not shrink by $(1-w/2)^{|S|}$, as \cref{ex:pathblur} shows.

\begin{example}[The path, smoothed]\label{ex:pathblur}
In \cref{ex:path} with $w=1/2$, the smoothed guide loses approximately $0.0449$ of linear score and gains $0.0018$ in the potential term.
Its objective value is approximately $\log\frac{13}8-0.0431$, against the loss bound $w^2\sum_i\E[c_i(1-c_i)]\approx0.0880$ in~\eqref{sm:loss}.
The exact interaction energies of row~$2$ are
\begin{center}
\begin{tabular}{lcccc}
\toprule
$S$ & $\varnothing$ & $\{1\}$ & $\{2\}$ & $\{1,2\}$ \\
\midrule
$\|(f_2)_S\|_2^2$ & $9/169$ & $8/2535$ & $2/13$ & $4/195$ \\
$\|(g_2)_S\|_2^2$ & $9/169$ & $2/845$ & $619/5070$ & $11/1690$ \\
\bottomrule
\end{tabular}
\end{center}
With $\rho=1-w/2=3/4$, the table gives
\[
\sum_{S\subseteq\{1,2\}}\rho^{-|S|}\|(g_2)_S\|_2^2 =\frac9{169}+\frac43\left(\frac2{845}+\frac{619}{5070}\right) +\frac{16}9\frac{11}{1690} =\frac3{13}=\|f_2\|_2^2.
\]
The interaction energy on $\{2\}$ is multiplied by $619/780$, which exceeds $\rho=3/4$.
\end{example}

More generally, fix a common $0<\rho\le1$ at every node.
If each threshold $\iota_c$ is replaced by a function of its input with mean $p=1-c$ and variance $\rho p(1-p)$, the recursion~\eqref{sm:recursion} gives the same identity with this $\rho$.
If every variance is at most $\rho p(1-p)$, it gives the corresponding inequality.
For example, resampling the queried input replaces $\iota_c$ by $(1-\tau)\iota_c+\tau(1-c)$ and turns $\mathcal T$ into $T_\tau \mathcal T$.
Its variance factor is $(1-\tau)^2$, and the identity holds component by component: $\sum_S(1-\tau)^{-2|S|}\|(T_\tau \mathcal T)_S\|_2^2=\|\mathcal T\|_2^2$.
With $(1-\tau)^2=1-w/2$, that is, $\tau=w/4+O(w^2)$, resampling removes the same fraction of variance as threshold smoothing, but on a fair bit it lowers the score by $\tau\log2$, whereas threshold smoothing lowers it only from $\log2$ to $\Hs(1/2+w/4)=\log2-O(w^2)$.

\paragraph{The cost of variance reduction.}
The energy identity shows how variance contraction at each comparison controls the decay of high-degree tails.
We now bound the linear-score loss required for a given variance reduction.

\begin{proposition}[The cost of smoothing one threshold]\label{sm:onecut}
For $c\in(0,1)$, put $p=1-c$ and recall $\iota_c=\ind_{[c,1)}$.
Write $\Var(\phi)$ for the variance of $\phi(U_1)$.
Every measurable $\phi:(0,1)\to[0,1]$ with $\int_0^1\phi=p$ satisfies
\begin{equation}\label{disc:onecut}
\Hs(p)-\int_0^1a\phi \ge\frac{3}{32\,c(1-c)} \bigl(\Var(\iota_c)-\Var(\phi)\bigr)^2.
\end{equation}
\end{proposition}

\begin{proof}
The map $\phi\mapsto1-\phi(1-\cdot)$ exchanges $c$ and $1-c$ and, since $a$ is antisymmetric with $\int_0^1a=0$, preserves both sides of~\eqref{disc:onecut}; so assume $c\le1/2$.
Put $\psi=\iota_c-\phi$ and $t=\int_0^1|\psi|$.
Then $\int_0^1\psi=0$, $|\psi|\le1$, and $\psi$ has the same sign as $u-c$.
The variance deficit is
\[
\Var(\iota_c)-\Var(\phi) =\int_0^1\phi(1-\phi)\in[0,t],
\]
because $\phi(1-\phi)\le|\psi|$ pointwise.
Since $\int_0^1\psi=0$, each of $(0,c)$ and $[c,1)$ carries mass $t/2$ of $|\psi|$, and $t/2\le c$.
Since $\int_c^1a(u)\,du=\Hs(p)$,
\[
\Hs(p)-\int_0^1a\phi =\int_0^1\bigl(a(u)-a(c)\bigr)\psi(u)\,du,
\]
where $a(c)-a(u)\ge\log(c/u)\ge(c-u)/c$ for $u<c$ and $a(u)-a(c)\ge\log(u/c)\ge(u-c)/u$ for $u>c$.
On each side, $|a(u)-a(c)|$ increases with the distance from $c$.
For this fixed $t$, the bathtub principle minimizes the integral subject to these sign and mass constraints by placing $|\psi|=1$ on $[c-t/2,c+t/2)$.
Hence
\begin{align*}
\Hs(p)-\int_0^1a\phi &\ge\int_{c-t/2}^c\frac{c-u}{c}\,du +\int_c^{c+t/2}\frac{u-c}{c+t/2}\,du =\frac{t^2}{8c}+\frac{t^2}{8(c+t/2)}\\
&\ge\frac{3t^2}{16c} \ge\frac{3t^2}{32c(1-c)} \ge\frac{3}{32c(1-c)} \bigl(\Var(\iota_c)-\Var(\phi)\bigr)^2.\qedhere
\end{align*}
\end{proof}

Threshold smoothing removes variance $wc(1-c)/2$ (\cref{sm:kernel}) and loses at most $w^2c(1-c)$ of linear score (\cref{sm:encoder}).
For the variance contraction $\rho=1-w/2$, its linear-score loss is therefore within a factor $128/3$ of the minimum over all replacements with the same mean and variance, uniformly for $0<c<1$.

\section{Closing remarks}\label{sec:latticescope}

\subsection{Discussion}\label{sec:scope}

\paragraph{Log-supermodularity.}
Our algorithm treats \BIS\ through the more general problem of counting with efficiently evaluable positive log-supermodular weights.
Log-supermodularity also plays a central role in classifications of approximate counting problems~\cite{BDGJM13,CDGJLMR15}.
The general log-supermodular counting problem is \BIS-hard, but is not known to be \BIS-equivalent.
Our algorithm may therefore be solving a strictly harder problem than \BIS.
Can one establish \BIS-equivalence, or identify a separation that can be exploited to obtain faster algorithms for \BIS?

\paragraph{Bipartite and general hard-core models.}
Extending our counting or sampling bound to general hard-core models, including fugacities approaching the uniqueness threshold, would also speed up SAT algorithms.
Our companion paper~\cite[Theorems~1.6 and~2.8]{JLY26} shows that an $\exp(\widetilde O(n^{1/3}))$ hard-core sampling algorithm for general graphs at fixed accuracy in this setting would give a randomized SAT algorithm running in $\exp(\widetilde O(s^{3/5}))$ time on inputs of binary length $s$.

\subsection{Open questions}\label{sec:openquestions}

\paragraph{Improving the running time.}
For arbitrary bipartite graphs, can the exponent $1/3$ be reduced, and is a quasi-polynomial-time approximation scheme or an FPRAS possible?
Within our approach, \cref{conv:conversion} reduces this question to the accuracy and cost of lower estimates on every pinned residual.
Additive error $\log^{O(1)}n$ at quasi-polynomial cost would suffice for a quasi-polynomial scheme; error $O(\log n)$ at polynomial cost would suffice for an FPRAS and polynomial-time sampling.

\paragraph{Improving the estimator.}
Write $Q=\log d\,\log(d+B)$.
For fixed $p>0$, an objective gap $O(dw^p)$ and estimation cost $\exp(O(Q/w))$ up to polynomial factors, uniformly over residuals and $0<w\le1/2$, give a total exponent $O(dw^p+Q/w)$ by \cref{conv:conversion}.
Balancing at $w=(Q/d)^{1/(p+1)}$, with enumeration if this exceeds $1/2$, gives
\[
O\left(d^{1/(p+1)}Q^{p/(p+1)}\right).
\]
Resampling has $p=1$ with $w=\tau$, and threshold smoothing has $p=2$.
Changing only the constants in these bounds leaves these powers unchanged.

\Cref{sm:onecut} shows that threshold smoothing is optimal up to an absolute constant in linear-score loss at its given variance reduction.
Potential gains can offset score losses (\cref{ex:pathblur}), so the proposition gives no corresponding lower bound on the full objective gap.
Can a sharper analysis of the full objective, a different guide construction, or better control of interaction tails improve this balance and reduce the exponent $1/3$?

\paragraph{Quantitative reductions.}
Approximation-preserving reductions may enlarge instances and alter accuracy requirements, so their quantitative costs must be composed with our running-time bound.
Which problems in the \BIS-equivalence class admit reductions, or direct log-supermodular representations, that preserve a comparable saving in their natural size parameters?
\section*{Acknowledgements}
We thank Weiming Feng and Heng Guo for constructive feedback during early stages of the project.

\section*{AI methodology}
The coarse estimator idea first emerged during interactions with GPT~5.5~Pro.
The first such construction was suggested by GPT~6~Pro on 7~October~2026, when the authors asked specifically for ideas that make use of convex optimization (with pointers to OpenAI's math GitHub repo) and ``the difference between bipartite $q$-coloring and \#BIS'' (with a note on bipartite $q$-coloring from Heng Guo through private communication).
The initial proof completed by AI claimed an algorithm running in time $\exp\left(n \left(\frac{(\log \log n)^2}{\log n}\right)^{1/5}\right)$.
We subsequently identified the Sinclair--Jerrum walk on the self-reduction tree, the variational principle on that tree, and the use of the Lov\'asz extension as implicit ingredients hidden in the proofs.
We also abstracted the log-supermodular structure underlying the algorithm, leading to the current organization and presentation of the paper.
OpenAI Codex assisted with mathematical development, finite-algorithm accounting, source drafting, literature checks, and bibliography preparation.
We also used Codex and Claude to help polish and check proofs.

\begingroup
\raggedright \interlinepenalty=10000
\bibliographystyle{alphaurl}
\bibliography{refs-bis}
\endgroup

\appendix
\crefalias{section}{appendix} \crefalias{subsection}{appendix}
\section{Algorithmic and bit implementation details}\label{sec:finite}

We first give the quantitative conversion promised in \cref{conv:conversion}, then implement the algorithm with bit operations and independent fair random bits.

\paragraph{Conventions.}
In the exact-arithmetic algorithm, a fixed rule resolves bin and transition boundaries.
Integer indices are obtained by comparisons and binary search, with the logarithm of their range charged to the operation count.
The bit implementation uses the following conventions.
Integers are encoded in binary, and a rational by its numerator and denominator.
Arithmetic, comparisons and memory access are charged by bit complexity.
Every loop, transition and evaluation has a predetermined cap, so all bounds hold on every execution.
There is no transcendental primitive: logarithms of rationals and the constant $\log2$ are computed to absolute error $2^{-\ell}$, with certified error bounds, in time polynomial in $\ell$ and the input length~\cite{BZ10}.
The evaluator returns $W(x)$ for every configuration $x$ as a rational whose numerator and denominator have at most $B$ bits, in at most $T_W$ bit operations.
Accuracy and failure-probability parameters are rationals, and $\ell_{\rm par}$ denotes the total length of their encodings.
Each is first rounded down to a power of two, within a factor two, at cost $\poly(\ell_{\rm par})$.

\begin{theorem}[Bit model]\label{thm:finite}
When the cost counts bit operations and fair random bits, \cref{thm:lsm,thm:main,alg:additive} hold for rational accuracy and failure-probability parameters with the same exponential factors and with the polynomial factors multiplied by $\poly(\ell_{\rm par})$.
On every execution the sampler of \cref{thm:lsm} returns a configuration, that of \cref{thm:main} an independent set, and the counters and additive estimator return rationals.
\end{theorem}

\subsection{Quantitative conversion from coarse estimates}\label{sec:conversiondetails}

Keep the tree and residual notation of \cref{sec:selfreduction}.
We give the tree walk and the failure-probability and time bounds used in \cref{conv:conversion}.
The conversion is classical~\cite{JVV86,SJ89}; we track the constants because the ratio $e^b$ between estimates and masses, the distortion of~\eqref{conv:distortion} below, is superpolynomial and its exponent enters the main bound, and because \cref{sec:finitealgorithm} needs explicit caps.
The walk is that of Sinclair and Jerrum~\cite[Section~4]{SJ89}, and \cref{w:tree} states the weighted form of their bounds that we use.

\paragraph{Counting from residual samples.}
\begin{proposition}[Counting from the sampler]\label{conv:counting}
Suppose that the residual law of every node can be sampled within total variation distance $\eta$ at cost $C_{\rm s}\poly(d,\log(1/\eta))$.
Then relative counting of $Z_W$ with error $0<\epsilon\le1/2$ and failure probability $0<\delta\le1/2$ costs $(C_{\rm s}+T_W)\poly(d,\epsilon^{-1},\log(1/\delta))$ on every execution.
\end{proposition}
\begin{proof}
Put $n_{\rm c}=\ceil{8192d^2\epsilon^{-2}}\ceil{\log_2(4d/\delta)}$.
For $i=1,\ldots,d$, with the prefix $x_{<i}$ already chosen, draw $n_{\rm c}$ samples from the residual law of $x_{<i}$ within total variation distance $\epsilon/(128d)$, each with fresh randomness.
Let $x_i$ be the more frequent value of coordinate $i$ among them, with ties broken toward zero, and let $\widehat\mu_i\ge1/2$ be its frequency.
Return $\widehat Z=W(x)/\prod_i\widehat\mu_i$.

With $\mu_i=\mu_W(X_i=x_i\mid X_{<i}=x_{<i})$, the chain rule~\eqref{conv:chain} gives $Z_W=W(x)/\prod_i\mu_i$ exactly.
Conditional on the earlier choices, each value of coordinate $i$ has a sampled frequency whose mean is within $\epsilon/(128d)$ of its conditional probability.
By Hoeffding's inequality~\cite{Hoe63}, a further deviation larger than $\epsilon/(128d)$ has probability at most $2\exp(-2n_{\rm c}\epsilon^2/(128d)^2)\le\delta/(2d)$.
The two frequencies are complementary, so with probability at least $1-\delta$ every level has $|\widehat\mu_i-\mu_i|\le\epsilon/(64d)$, and then $\mu_i\ge1/2-1/128\ge1/3$.
The logarithm is $3$-Lipschitz on $[1/3,1]$, so
\[
\Bigl|\log\frac{\widehat Z}{Z_W}\Bigr| \le3\sum_i|\widehat\mu_i-\mu_i|\le\frac{3\epsilon}{64},
\]
and $1-\epsilon\le\widehat Z/Z_W\le1+\epsilon$ for $0<\epsilon\le1/2$.
On every execution $\widehat\mu_i\ge1/2$, so $\widehat Z$ is finite.
The algorithm makes exactly $dn_{\rm c}$ sampler calls, each of cost $C_{\rm s}\poly(d,\log(d/\epsilon))$, and one evaluation of $W$.
\end{proof}

\paragraph{The Sinclair--Jerrum walk.}\label{sec:sjwalk}

As explained in \cref{sec:coarseconversion}, Sinclair and Jerrum~\cite[Section~4]{SJ89} replace the one-way descent by a reversible walk on the tree.
Suppose that every node $v$ carries a mass estimate $q_v>0$ and that, for a \emph{distortion} $\Bd\ge1$,
\begin{equation}\label{conv:distortion}
Z_v/\Bd\le q_v\le Z_v\quad\text{at every node }v,\qquad q_x=W(x)\quad\text{at every leaf }x.
\end{equation}
No estimate is required to equal the sum of the estimates of its children.
For $\depth(v)<d$ the estimates determine the conditional marginal probabilities only within the factor $\Bd$, $p_v/\Bd\le q_{v1}/(q_{v0}+q_{v1})\le\Bd p_v$, so a descent driven by them loses a factor at every level.
The walk avoids this loss.

As in~\cite[(8)--(10)]{SJ89}, give the edge between a node $v\ne o$ and its parent $v^-$ the \emph{weight} $q_v$, the estimate at its lower endpoint, and give every node $v$ the \emph{weighted degree}
\begin{equation}\label{w:tree-notation}
s_v=\ind\{v\ne o\}\,q_v+\sum_{y\text{ child of }v}q_y,
\end{equation}
the total weight of its edges, which Sinclair and Jerrum write $d(v)$.
Let $S_o=\sum_vs_v$.
The walk $P_{\rm tr}$ holds with probability $1/2$ and otherwise moves along an edge at the current node, chosen with probability proportional to its weight:
\[
P_{\rm tr}(v,y)=\frac{q_y}{2s_v}\quad(y\text{ a child of }v),\qquad P_{\rm tr}(v,v^-)=\frac{q_v}{2s_v}\quad(v\ne o),\qquad P_{\rm tr}(v,v)=\frac12 .
\]
Exact leaf weights make its stationary law, conditioned on the leaves, equal to $\mu_W$; the distortion affects only the mixing time and the stationary probability of the leaves.

\begin{lemma}[Sinclair--Jerrum walk]\label{w:tree}
Consider any rooted tree whose leaves all have depth $d\ge1$, with the chain $P_{\rm tr}$ and the notation above defined on it.
Give each leaf $x$ a weight $\omega_x>0$, let $Z_v$ be the sum of the weights of the leaves below $v$, and suppose that $Z_v/\Bd\le q_v\le Z_v$ at every node and $q_x=\omega_x$ at every leaf.
Put $h_0=2\Bd d$ and $\Gamma=32\Bd^4d^2$, and let $\mathcal L$ be the set of leaves.
Then $P_{\rm tr}$ is reversible with stationary law $\pi(v)=s_v/S_o$; moreover $\pi(\mathcal L)\ge1/h_0$, $\pi(o)\ge1/(h_0\Bd)$, $\pi(x\mid\mathcal L)=\omega_x/Z_o$ at each leaf $x$, and, for $0<\eta_0<1$,
\begin{equation}\label{w:mixing}
\|P_{\rm tr}^t(o,\cdot)-\pi\|_{\TV}\le\eta_0 \quad\text{whenever}\quad t\ge\Gamma\log\frac{h_0\Bd}{\eta_0}.
\end{equation}
\end{lemma}
\begin{proof}
Every estimate is within the ratio $r=\Bd$ of its mass, and every leaf carries its exact weight.
The proofs of~\cite[(11), Lemma~4.1 and Lemma~4.2]{SJ89} use only these two facts and the disjointness of the subtrees at each depth, so they hold verbatim with the number of leaves below a node replaced by its mass and with their depth $m$ equal to our $d$.
They give the stationary law, $\pi(\mathcal L)\ge1/(2rd)=1/h_0$, and conductance at least $1/(4r^2d)$, half of whose square is $1/\Gamma$.
The conditional leaf law follows from $s_x=q_x=\omega_x$.
By~\cite[(12)]{SJ89} with leaf weights, $S_o\le2rdZ_o$, while $s_o=\sum_yq_y\ge Z_o/\Bd$, the sum over the children $y$ of the root; hence $\pi(o)\ge1/(2r^2d)=1/(h_0\Bd)$.
Finally, the chain holds with probability $1/2$, and it is ergodic because every $q_v$ is positive, so~\cite[Theorem~3.4]{SJ89}, applied with the set $\{o\}$ in the role of their $U$, gives $P_{\rm tr}^{2t}(o,o)/\pi(o)-1\le(1-1/\Gamma)^{2t}/\pi(o)$.
By reversibility, $\sum_v(P_{\rm tr}^t(o,v)-\pi(v))^2/\pi(v)=P_{\rm tr}^{2t}(o,o)/\pi(o)-1$, so by the Cauchy--Schwarz inequality $2\|P_{\rm tr}^t(o,\cdot)-\pi\|_{\TV}\le(1-1/\Gamma)^t\pi(o)^{-1/2} \le(h_0\Bd)^{1/2}e^{-t/\Gamma}$, which proves~\eqref{w:mixing}.
\end{proof}

\paragraph{Additive estimates and a cached table.}\label{sec:globalguide}

Assume the estimator guarantee~\eqref{conv:oracle} of \cref{sec:coarseconversion}, with the same additive error bound $b$ for all nodes.

\begin{proposition}[Sampling from additive estimates]\label{conv:sampling}
Given~\eqref{conv:oracle}, for every $0<\eta\le1/2$ there is a sampler within total variation distance $\eta$ of $\mu_W$, and one within $\eta$ of the residual law $W_v/Z_v$ of every node $v$.
Every execution uses at most $\exp(O(b+1))\,C_0\poly(d,\log(1/\eta))$ operations and uniform draws.
\end{proposition}

Each sampler invocation, including each call in \cref{conv:counting}, uses fresh randomness and an initially empty cache.
Within an invocation, cache the first answer of the estimator for each node $v$ and set
\begin{equation}\label{conv:B}
q_v=W_v(0)\,e^{A_v},\qquad \Bd=e^b .
\end{equation}
At a leaf $x$ the estimate is the exact weight $q_x=W(x)$.
On the success event in~\eqref{conv:oracle} these estimates satisfy~\eqref{conv:distortion}, and on every execution, including a failure of the estimator, they are positive.

For the analysis, assign an independent seed of the estimator to each of the fewer than $2^{d+1}$ nodes, with $\delta_{\rm est}=\eta/2^{d+5}$; then with probability at least $1-\eta/16$ all estimates satisfy~\eqref{conv:distortion}.
The algorithm builds this table lazily: it generates a seed and stores the resulting estimate on the first request for a node, as in the implementation that remembers earlier answers~\cite[Remark after Theorem~4.5, pp.~118--119]{SJ89}.
The seed for each new node is independent of all previous randomness, even when requests are adaptive, so deferred generation has the same law as the complete table.
Since $\log(1/\delta_{\rm est})=O(d+\log(1/\eta))$, each request costs $C_0\poly(d,\log(1/\eta))$.

\begin{proof}[Proof of \cref{conv:sampling}]
Use the cached mass estimates~\eqref{conv:B} and the chain of \cref{w:tree} with leaf weights $\omega_x=W(x)$, and choose integer caps
\[
t_{\rm walk}=\ceil{\Gamma\log\frac{32h_0^2\Bd}{\eta}}, \qquad n_{\rm ch}=\ceil{2h_0\log\frac{16}{\eta}}, \qquad n_{\rm tr}=n_{\rm ch}t_{\rm walk}.
\]
Run $n_{\rm ch}$ independent chains from the root for $t_{\rm walk}$ steps each, sharing the same cache and using fresh transition randomness.
Return the configuration of the first endpoint that is a leaf, and the all-zero configuration if none is a leaf.

Condition on a table for which~\eqref{conv:distortion} holds.
By \cref{w:tree} the endpoint law $\pi'=P_{\rm tr}^{t_{\rm walk}}(o,\cdot)$ of every chain is within $\eta_0=\eta/(32h_0)$ of $\pi$.
Each chain ends at a leaf with probability at least $1/h_0-\eta_0\ge1/(2h_0)$, and
\[
\|\pi'(\,\cdot\mid\mathcal L)-\pi(\,\cdot\mid\mathcal L)\|_{\TV} \le \frac{2\eta_0}{1/h_0-\eta_0}<\eta/8 .
\]
The first leaf endpoint has the conditional law $\pi'(\cdot\mid\mathcal L)$.
The probability of no leaf endpoint is at most $e^{-n_{\rm ch}/(2h_0)}\le\eta/16$, and a bad table adds at most $\eta/16$, so the total variation error is smaller than $\eta$.
Each transition is implemented exactly: one uniform draw selects the hold event and, if needed, an edge by cumulative comparison of the positive edge weights.
For the residual law of a node $v$ of depth less than $d$, run the same construction on the subtree below $v$.
\Cref{w:tree} applies to it with leaves at depth $d-\depth(v)$.
Its constants $h_0$ and $\Gamma$ are no larger than those of the whole tree, so the same caps suffice.
At a leaf the residual law is a point mass, returned exactly.

There are $n_{\rm tr}=\exp(O(b+1))\poly(d,\log(1/\eta))$ transitions and at most $3n_{\rm tr}+1$ new requests, each costing $C_0\poly(d,\log(1/\eta))$.
Mass estimates are stored in a binary trie keyed by the prefixes, and the current path is kept on a stack, so lookups and updates take $\poly(d)$ operations.
The integer caps are found by doubling and binary search in $O(\log(n_{\rm tr}+2))$ operations.
These bounds hold on every execution, regardless of failures of the estimator.
\end{proof}

Together, \cref{conv:sampling,conv:counting} prove \cref{conv:conversion}.

\subsection{Finite-precision arithmetic and bounded random choices}\label{sec:finitechoices}

Two standard lemmas replace exact uniform draws.

\begin{lemma}[Accumulated simulation error]\label{finite:simulation}
Consider a randomized procedure with at most $t$ sequential steps.
If, conditional on every common past, the ideal and the implemented step $j$ have output laws at total variation distance at most $\eta_j$, then the final outputs are at total variation distance at most $\sum_{j\le t}\eta_j$.
\end{lemma}
\begin{proof}
Couple each pair of steps maximally while the histories agree.
By a union bound, the histories disagree with probability at most $\sum_{j\le t}\eta_j$, and otherwise the outputs coincide.
\end{proof}

\begin{lemma}[Finite choices from fair bits]\label{finite:choice}
Let $\omega_1,\ldots,\omega_{n_\omega}$ be positive rationals whose numerators and denominators have at most $\ell_\omega$ bits, and let $\ell\ge1$.
Using $\ell$ fair bits and $\poly(n_\omega,\ell_\omega,\ell)$ bit operations one can draw an index whose law is within total variation distance $(n_\omega-1)2^{-\ell}$ of $(\omega_j/\sum_i\omega_i)_j$.
\end{lemma}
\begin{proof}
Read the bits as a grid point $y\in\{0,2^{-\ell},\ldots,1-2^{-\ell}\}$ and return the least $j$ with $y<\sum_{i\le j}\omega_i/\sum_i\omega_i$, by exact integer cross multiplication.
For $U$ uniform on $[y,y+2^{-\ell})$, this agrees with the inverse transform of $U$ unless the cell contains one of the $n_\omega-1$ interior cumulative boundaries.
\end{proof}

Without truncation, fair bits generate such choices exactly with a random number of bits~\cite{KY76}; the truncation makes the cost bounded on every execution.

\subsection{Finite-precision optimization and simulation}\label{sec:finitealgorithm}

\paragraph{The estimator.}
At a node whose residual has $k\ge1$ free coordinates, run the generic estimator of \cref{alg:generic} with $\gamma=1/128$ in place of $1/64$.
Choose a known bound $b_*$ on the objective gap, and choose a degree $D$ after the rounded parameter $\kappa$ of item~(b) below has been computed.
For threshold smoothing use $b_*=dw^2/4$ and $D=\min\{d,\ceil{4\kappa/w}\}$; for ordinary resampling use the rational bound $b_*=d\tau$ and $D=\min\{d,\ceil{\kappa/\tau}\}$.
These choices satisfy the required tail bound with the rounded $G_0$ of item~(b); if $D=d$, the tail vanishes.
Make the following further changes.
\begin{enumerate}[label=(\alph*)]
\item
Each bin $Y_i$ is read from $2s$ fair bits, since $m=4^s$.
\item
Each bin score $\bar a_j$ is replaced by a dyadic rational $\tilde a_j$ with $0\le\bar a_j-\tilde a_j\le1/(64d)$, computed from certified values of $\Hs(j/m)$.
In the definitions of $G_0$ and $\kappa$ in~\eqref{alg:parameters}, $A_0$ is replaced by the integer $2s+2\ge A_0+1$, so that $G_0=2s+2+4B$; in~\eqref{alg:validation}, $\log(8n_{\rm att}/\delta)$ is replaced by the larger integer $\ceil{\log_2(8n_{\rm att}/\delta)}$.
All parameters are then integers or rationals computed exactly.
\item
The residual weight $W_v(x')/W_v(0)$ is formed exactly, a rational whose numerator and denominator have at most $2B$ bits.
Each value $\Phi(S_t)$, $t\ge1$, entering the potential term is computed to absolute error $\xi=\gamma/(64P)$ on a common dyadic grid, and $\Phi(\varnothing)=0$ is exact.
\item
Let $\alpha$ be the power of two with $\gamma/(2MG_1^2)<\alpha\le\gamma/(MG_1^2)$, take $n_{\rm it}=\ceil{2PMG_1^2/\gamma^2}$ steps, and return $\max_\rho(\widehat J^{(\rho)}-\gamma-3P\xi)$, clamped as before, where $\widehat J^{(\rho)}$ averages the computed sample values.
\end{enumerate}
Soundness is unchanged: for the clipped guide $F=\clip(u)$ of the proof of \cref{alg:soundness}, $\psi_{\tilde a_j}(u_i)\le\tilde a_jF_i\le\bar a_jF_i$ because $F_i\ge0$, so the objective with scores $\tilde a_j$ is at most $\log Z_\Phi$ for every candidate.
In the proof of \cref{alg:witness}, the rounded scores lower the objective at $\bar g$, which takes values in $[0,1]^k$, by at most $k/(64d)\le1/64$.
The truncation step is unchanged, because $\kappa$ is defined from the new $G_0$, which bounds $|\tilde a_j|+2\Lambda$.
The losses are $1/64$ from score rounding, $1/64$ from the bins, and $1/128$ from truncation, totaling $5/128$.
Thus $J(\theta^\circ)\ge\log Z_\Phi-b_*-5/128$.

In the analysis below, $J(\theta)$ denotes the objective with the dyadic scores $\tilde a_j$ and the exact potential $\Phi$.
Write the potential term as $\PL(z)=\sum_{t=1}^k\Phi(S_t)(z_{j_t}-z_{j_{t+1}})$, where $z_{j_1}\ge\cdots\ge z_{j_k}$ and $z_{j_{k+1}}=0$.
The first $k-1$ differences are nonnegative, and
\[
\sum_{t=1}^{k-1}|z_{j_t}-z_{j_{t+1}}|+|z_{j_k}| =z_{j_1}-z_{j_k}+|z_{j_k}| \le3\max_i|z_i|.
\]
Thus errors of at most $\xi$ in the values $\Phi(S_t)$ change the computed sample objective at $z=u^\theta(Y)$ by at most $3\xi\max_i|z_i|\le3P\xi$, and each chain increment by at most $2\xi$.

Rounded values need not define a supermodular potential.
For a fixed sorted chain, let $\zeta_Y(\theta)$ be the supergradient obtained from the exact potential increments and let $\widetilde\zeta_Y(\theta)$ be the computed vector.
A computed row slope has magnitude at most
\[
|\tilde a_j|+2\Lambda+2\xi\le G_0,
\]
so the bounds~\eqref{alg:samplebounds} remain valid, and
\[
\|\widetilde\zeta_Y(\theta)-\zeta_Y(\theta)\|_2 \le2\xi\sqrt P.
\]
Since $\|\theta-\theta^*\|_2\le2\sqrt P$, the additional term in the distance recursion contributes at most $4P\xi=\gamma/16$ to~\eqref{alg:optimizerbound}.
Consequently
\[
\E[J(\theta^*)-J(\bar\theta)]\le\frac{17\gamma}{16},
\]
and Markov's inequality gives $J(\bar\theta)\ge J(\theta^*)-17\gamma/4$ with probability at least $3/4$.

Validate as in~\eqref{alg:validation}.
Let $\widehat J_{\rm true}^{(\rho)}$ be the validation average using the exact potential values in attempt $\rho$.
The computed average satisfies
\[
|\widehat J^{(\rho)}-\widehat J_{\rm true}^{(\rho)}| \le3P\xi .
\]
With probability at least $1-\delta/4$, every $\widehat J_{\rm true}^{(\rho)}$ differs from $J(\bar\theta^{(\rho)})$ by at most $\gamma$.
On this event every corrected value $\widehat J^{(\rho)}-\gamma-3P\xi$ is at most $J(\bar\theta^{(\rho)})\le\log Z_\Phi$.

With probability at least $1-4^{-n_{\rm att}}\ge1-\delta/4$, some ascent attempt satisfies $J(\bar\theta^{(\rho)})\ge J(\theta^*)-17\gamma/4$.
On the intersection of the two events, the largest corrected validation value is at least
\[
\log Z_\Phi-b_*-\frac5{128} -\Bigl(\frac{17}4+2\Bigr)\gamma-6P\xi \ge\log Z_\Phi-b_*-\frac18 .
\]
The logarithms need $O(\log P)$ bits of precision beyond the input length, the iterates stay on a common dyadic grid in $[-1,1]$, and the cost is~\eqref{alg:genericcost}, now in bit operations and fair bits.

\paragraph{Additive estimation.}
To implement \cref{alg:additive}, let $\epsilon_0$ be the rounded dyadic accuracy, with $\epsilon/2<\epsilon_0\le\epsilon$, and choose the largest power of two $w$ with $w^2\le\epsilon_0$.
Integer comparisons find it, and $\sqrt{\epsilon_0}/2<w\le\sqrt{\epsilon_0}\le1/4$.
Run the preceding estimator at the root with $b_*=dw^2/4$ and the rounded failure probability, and compute a rational $L_0$ satisfying $0\le\log W(0)-L_0\le\epsilon_0/4$ by certified logarithm evaluation.
Return the sum $A$ of $L_0$ and the estimate at the root.
With probability at least $1-\delta$,
\[
0\le\log Z_W-A \le\frac{dw^2}{4}+\frac18+\frac{\epsilon_0}{4} \le\frac{\epsilon_0d}{2}+\frac18 \le\epsilon d+\frac18.
\]
Since $w^{-1}=O(\epsilon^{-1/2})$, the exponential factor is unchanged; the additional logarithm uses $\poly(B,\ell_{\rm par})$ bit operations.

\paragraph{The walk.}
Let $A_{\max}=d(1+2B)$, so that $0\le A_v\le A_{\max}$, compute a rational $\widetilde{\log2}$ with $\log2\le\widetilde{\log2}\le\log2+1/(2(A_{\max}+1))$, and use the mass estimates
\[
q_v=W_v(0)\,2^{\floor{A_v/\widetilde{\log2}}},
\]
and $q_x=W(x)$ at leaves.
Then $0\le A_v-\floor{A_v/\widetilde{\log2}}\log2 <A_v(\widetilde{\log2}-\log2)/\widetilde{\log2}+\log2<2$, so on the success event $e^{-b-2}Z_v\le q_v\le Z_v$, and on every execution the estimates are positive rationals of polynomial length.
\Cref{w:tree} applies with the power of two $\Bd=2^{\ceil{2(b+2)}}$.
With $\eta=2^{-\ell_\eta}$, take the integer caps $t_{\rm walk}=\Gamma\bigl(\ell_\eta+\ceil{\log_2(32h_0^2\Bd)}\bigr)$, $n_{\rm ch}=2h_0(\ell_\eta+4)$ and $n_{\rm tr}=n_{\rm ch}t_{\rm walk}$, and the estimator failure probability $\delta_{\rm est}=2^{-\ell_\eta-d-5}$.
Each transition draws the hold event with one fair bit and an edge among at most three by \cref{finite:choice} with $\ell=\ell_\eta+\ceil{\log_2n_{\rm tr}}+4$ bits.
Each step then has error at most $2^{1-\ell}\le\eta/(8n_{\rm tr})$, so by \cref{finite:simulation} the bit model adds at most $\eta/8$ to the error budget in the proof of \cref{conv:sampling}, and the total stays below $\eta$.

\paragraph{Counting.}
\Cref{conv:counting} calls this sampler at accuracy $\epsilon/(128d)$, rounded down to a power of two, and computes $n_{\rm c}$ with integer logarithms.
The frequencies are rationals with denominator $n_{\rm c}$ and are at least $1/2$, so $W(x)/\prod_i\widehat\mu_i$ is a rational of polynomial length on every execution.

\begin{proof}[Proof of \cref{thm:finite}]
The additive-estimation construction above proves \cref{alg:additive} in the bit model.
For \cref{thm:lsm}, the balance is unchanged, with $\ell_d$ and $\ell_B$ as in its proof: $w$ is a power of two found by integer comparisons, and the degree $D=\min\{d,\ceil{4\kappa/w}\}$ makes each estimator call cost
\[
\exp(O(w^{-1}\ell_d\ell_B))\poly(d,B,T_W,\log(1/\delta_{\rm est}))
\]
bit operations and fair bits.
The ordinary-resampling choice likewise gives the square-root bound of \cref{alg:baseline} in the bit model: take the power of two $\tau$ of that proof, give the walk the rational additive error bound $b=d\tau+1/8\ge d\tau\log2+1/8$, and use the corresponding degree $D$.
In the enumeration case the $2^d$ weights are summed exactly, with $\exp(O(d))\poly(B)$ bits, and a configuration is drawn by \cref{finite:choice} with $n_\omega=2^d$ and $\ell=d+\ell_\eta+1$, within $\eta/2$.
For \cref{thm:main}, the weights~\eqref{intro:shoreweight} are integers with at most $n+1$ bits, and the completion by a uniform subset of $R\setminus N(S)$ uses at most $n$ fair bits.
\end{proof}

\end{document}